\documentclass[final]{IEEEtran}
\usepackage{algorithm}
\usepackage{color}
\usepackage{algpseudocode}
\usepackage{graphics}
\usepackage{epsfig}
\usepackage{cite}
\usepackage{threeparttable}
\usepackage{graphicx}
\usepackage{picinpar}
\usepackage{amssymb}
\usepackage[cmex10]{amsmath}

\usepackage{bm}
\usepackage{setspace}
\usepackage{subfigure}
\usepackage{url}

\input epsf

\begin{document}

\title{\huge{Waveform Design for Joint Sensing and Communications in Millimeter-Wave and Low Terahertz Bands}}

\author{Tianqi Mao, \IEEEmembership{Member,~IEEE}, Jiaxuan Chen, Qi Wang, Chong Han, \IEEEmembership{Member,~IEEE}, Zhaocheng Wang, \IEEEmembership{Fellow,~IEEE}, and  George K. Karagiannidis, \IEEEmembership{Fellow,~IEEE}
\thanks{This work was supported in part by the National Key R\&D Program of China under Grant 2018YFB1801501,
in part by Shenzhen Special Projects for the Development of Strategic Emerging Industries (201806081439290640),
and in part by Shenzhen Wireless over VLC Technology Engineering Lab Promotion. \emph{(Corresponding author: Zhaocheng Wang.)}}
\thanks{T. Mao is with the School of Electronic and Information Engineering, Beihang University, Beijing 100191, China (e-mail: maotq@buaa.edu.cn).}
\thanks{Z. Wang is with Beijing National Research Center for Information Science and Technology, Department of Electronic Engineering, Tsinghua University, Beijing 100084, China, and Z. Wang is also with Shenzhen International Graduate School, Tsinghua University, Shenzhen 518055, China. (e-mail: zcwang@tsinghua.edu.cn).} %
\thanks{J. Chen is with Huawei Technologies Co. Ltd., Shenzhen 518129, China (e-mail: chenjiaxuan16@hotmail.com).}
\thanks{Q. Wang is with Huawei Device Co. Ltd., Shenzhen 518129, China (e-mail: steven\_wq@hotmail.com).}
\thanks{Chong Han is with the UM-SJTU Joint Institute, Shanghai Jiao Tong University, Shanghai 200240, China (e-mail: chong.han@sjtu.edu.cn).}
\thanks{G. K. Karagiannidis is with the Wireless Communications Systems Group (WCSG), Aristotle University of Thessaloniki, Thessaloniki 54 124, Greece (e-mail: geokarag@auth.gr).}
\vspace*{-5mm}} %

\maketitle
\begin{abstract}
The convergence of radar sensing and communication applications in the millimeter-wave (mmWave) and low terahertz (THz) bands has been envisioned as a promising technology, since it incorporates high-rate data transmission of hundreds of gigabits per second (Gbps) and mm-level radar sensing in a spectrum- and cost-efficient manner, by sharing both the frequency and hardware resources. However, the joint radar sensing and communication (JRC) system faces considerable challenges in the mmWave and low-THz scale, due to the peculiarities of the propagation channel and radio-frequency (RF) front ends. To this end, the waveform design for the JRC systems in mmWave and low-THz bands with ultra-broad bandwidth is investigated in this paper. Firstly, by considering the JRC design based on the co-existence concept, where both functions operate in a time-domain duplex (TDD) manner, a novel multi-subband quasi-perfect (MS-QP) sequence, composed of multiple perfect subsequences on different subbands, is proposed for target sensing, which achieves accurate target ranging and velocity estimation, whilst only requiring cost-efficient low-rate analog-to-digital converters (A/Ds) for sequence detection. Furthermore, the root index of each perfect subsequence is designed to eliminate the influence of strong Doppler shift on radar sensing. Finally, a data-embedded MS-QP (DE-MS-QP) waveform is constructed through time-domain extension of the MS-QP sequence, generating null frequency points on each subband for data transmission. Unlike the co-existence-based JRC system in TDD manner, the proposed DE-MS-QP waveform enables simultaneous interference-free sensing and communication, whilst inheriting all the merits from MS-QP sequences. Numerical results validate the superiority of the proposed waveforms regarding the communication and sensing performances, hardware cost as well as flexibility of the resource allocation between the dual functions.
\end{abstract}
\vspace{-3mm}

\section{Introduction}\label{S1}
Millimeter-wave (mmWave) communication has been regarded as a critical role in the fifth-generation (5G) network to alleviate the scarcity of frequency resources caused by the escalating mobile data traffic \cite{Hemadeh_survey_18,Jiang_ojcs_21}, which is further expanded to the low terahertz (THz) band over $100$ GHz to support ultra-fast data transfer on the order of 100 Gigabit per second (Gbps) for future sixth-generation (6G) networks \cite{Rappaport_survey_19,zcwang_mag_11,Saad_network_20}. The ultra-broad communication bandwidth available in the mmWave and low-THz bands enables a large number of bandwidth-consuming services, including augment/virtual reality, high-definition video communication and wireless backhauling, etc. \cite{Bailey_access_20,Xiao_jsac_17,Ian_PC_14,PC_survey_19}. Besides, secure data transmission can be realized for military use, thanks to the ``pencil''-like extremely narrow beam pattern in mmWave and THz communication systems\cite{Xiao_jsac_17,Ian_PC_14}. Except for their merits in wireless communication, the radiated signals also enable accurate sensing applications with ultra-high-resolution target range and velocity estimation \cite{Liu_tcom_20,Alouini_survey_2020}, attributed to huge operation bandwidth and high carrier frequency on the order of 100 GHz. Moreover, the narrow beamwidth of the radar signals could effectively mitigate the clutter effects caused by multi-path propagation, further enhancing the sensing performance.
\subsection{State-of the-Art}
As the transceiver hardware architectures for wireless communication and radar systems have become more and more similar, thanks to the advanced digital signal processing \cite{Sturm_IEEEproc_11}, it is viable to incorporate both systems together by sharing one single transceiver platform and the operating spectrum, leading to cost-efficient and compact hardware design, as well as improved spectral efficiency. This so-called joint radar sensing and communication (JRC) concept has been extensively investigated for lower-frequency bands in the literature. Authors in \cite{Sturm_IEEEproc_11,Chiriyath_tccn_17,Feng_survey_20} provide comprehensive overviews of the existing JRC techniques, which classify their principles into co-existence and co-design philosophies. For co-existence methods, both the radar and communication subsystems are treated as interferers to each other. A simple strategy which could fully mitigate the interference is to perform target sensing and data transmission in a time-division duplex (TDD) manner \cite{Zatman_conf_16}. However, the dual functions are sometimes required to be steadily available, which necessitates simultaneous operation of radar sensing and communication, causing performance degradation of the JRC systems due to inevitable mutual interference \cite{Zhou_taes_19}. Under such scenario, a joint precoder-decoder design was proposed to maximize the signal-to-interference-plus-noise ratio (SINR) for interference mitigation of the JRC systems \cite{Cui_conf_17}. Besides, the interference to the communication subsystem from radar sensing is reduced by solving a non-convex joint interference removal and data demodulation problem, where two optimization algorithms are invoked \cite{Li_twc_19}. More details on the interference cancellation of co-existing JRC systems can be found in \cite{Zheng_spm_19}. These strategies, however, induce additional computational complexity.

Alternatively, the co-design implementations of JRC systems were developed by employing integrated waveforms, that enable both radar sensing and communication. {In \cite{Sturm_vtc_09,Braun_conf_10,Liu_cl_17}, the orthogonal frequency division multiplexing (OFDM) waveform was applied to the JRC systems due to the merits such as capability of high-rate data transmission, one-tap equalization and robustness to multi-path effects, etc. Explicitly, in \cite{Sturm_vtc_09}, an enhanced method for the radar range profile calculation was developed based on frequency-domain channel estimation, which enhanced the dynamic range of radar sensing. A maximum-likelihood (ML) estimator of target range and velocity was also proposed based on the channel estimation methods \cite{Braun_conf_10}. Besides, in \cite{Liu_cl_17}, the power allocation strategy was optimized for the OFDM JRC waveform based on information theory. Despite its advantages, the OFDM signal inherently suffers from high peak-to-average power ratio (PAPR), causing undesirable clipping distortions originated from the nonlinearity of high power amplifiers. Moreover, the auto-correlation property\footnote{In this paper, auto-correlation property under zero Doppler shift is mainly concerned \cite{Sturm_IEEEproc_11,Kumari_tvt_18}. However, we should be aware that strong Doppler shift could degrade the peak-to-sidelobe level in the correlation process of radar sensing.} of OFDM signals cannot be ensured for quadrature amplitude modulation (QAM) constellations, causing possible range sidelobes that affect the sensing performance \cite{Zhou_taes_19,Popovic_tit_18}.} On the other hand, the single-carrier approach is capable of maintaining low PAPR values, which is suitable for mmWave and low-THz-band systems equipped with imperfect nonlinear devices \cite{Chung_tmtt_18,Ramadan_access_18}. In \cite{Zeng_tvt_20}, a cyclic-prefixed single-carrier (CP-SC) JRC system was investigated, where a low-complexity estimator for target range and velocity was proposed based on cyclic correlations and fast fourier transform (FFT) operations. However, the simple communication waveform employing quadrature phase shift keying (QPSK) was directly applied for sensing, whose auto-correlation property is imperfect due to randomness of the transmitted symbols. Moreover, \cite{Tang_wcl_19} combined PSK modulation and the direct sequence spread spectrum (DSSS) technique to generate an integrated waveform for single-carrier JRC systems, which could achieve sufficiently high peak sidelobe level via optimization methods, but suffers from inherent data rate loss.

\begin{figure*}[t!]
	\begin{center}
		\includegraphics[width=0.8\linewidth, keepaspectratio]{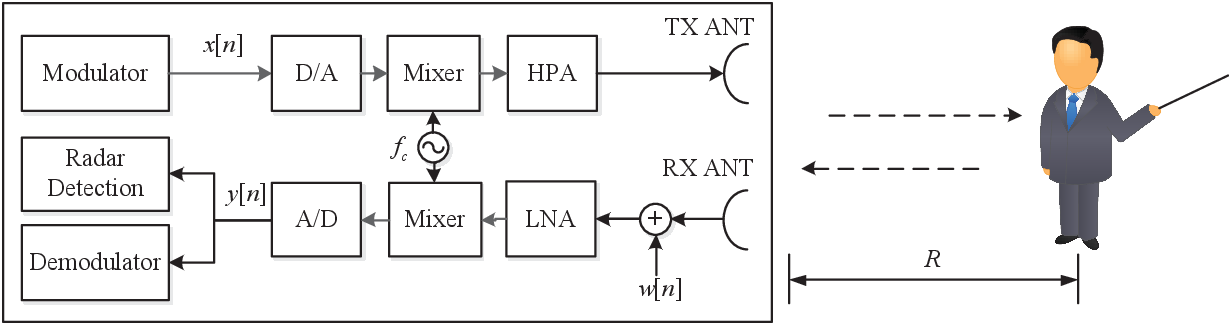}
	\end{center}
	\caption{Transceiver diagram of the JRC system at mmWave and low-THz frequencies.}
	\label{fig1}
	
\end{figure*}
\subsection{Motivation and Contributions}
The convergence of sensing and communications across the mmWave and low-THz frequencies are capable of enabling high-resolution target sensing, whilst attaining ultra-high-rate data transmission at the same time \cite{Petrov_wc_19}. However, there are significant technical challenges for its practical implementation, which cannot be fully overcome by the aforementioned schemes for lower-frequency bands. Firstly, in the mmWave and low-THz bands, the emitted signals are prone to severe path loss induced by the spreading loss and molecular absorption effects \cite{Hemadeh_survey_18,Lin_magazine_16}, leading to extremely low signal-to-noise ratio (SNR) at the radar receiver. Secondly, the Doppler shift becomes more dominant due to the ultra-high carrier frequency, which has negative impacts on the target sensing performance, e.g., false alarms caused by the range sidelobes \cite{Sturm_IEEEproc_11}. Besides, analog-to-digital converters (A/D) with sampling rate over tens of GHz may be required for mmWave/THz-JRC systems with ultra-broad bandwidth, which significantly enlarges the hardware cost. Finally, mutual interference between the dual JRC functions can be detrimental for their performances, especially for radar sensing, since the echo signals experiencing severe path loss could be easily submerged in the leaked communication signals to the radar receiver. At present, there have been preliminary researches concerning mmWave/low-THz JRC applications \cite{Dokhanchi_taes_19,Kumari_tvt_18,Duggal_taes_20,Kumari_tsp_20,Kumari_jstsp_21,Elbir_jstsp_21,Kumari_conf_20,Kumari_ojvt_21}. {Dokhanchi \emph{et al.} investigated the feasibility of the phase-modulated-continuous-wave (PMCW) and the orthogonal-frequency-division-multiple-access (OFDMA) waveforms in mmWave JRC systems with a bi-static radar \cite{Dokhanchi_taes_19}. {Despite their attractive merits, the PMCW waveforms inherently suffers from low data rate, whilst OFDMA signals have to deal with high PAPR issue as well as undesirable sidelobe level.} Besides, in \cite{Kumari_tvt_18}, the preamble of the IEEE 802.11 ad frame was employed for mmWave radar applications due to its good correlation property. The preamble sequence was further designed by incorporating Prouhet-Thue-Morse sequences to enhance the tolerance of Doppler shift effects \cite{Duggal_taes_20}. To enhance the estimation accuracy of target velocity of the preamble-based JRC system, Kumari \emph{et al.} proposed a virtual waveform design based on virtual preamble construction \cite{Kumari_tsp_20}. {These preamble-based JRC schemes, however, might be fragile to extremely severe path loss especially in the THz scale, due to their limited sequence length.} Moreover, the beamforming issues of the JRC systems at both the mmWave and THz scale were respectively discussed in \cite{Kumari_jstsp_21} and \cite{Elbir_jstsp_21}, where a desirable trade-off between the communication and radar sensing performances was achieved. In addition, \cite{Kumari_conf_20} and \cite{Kumari_ojvt_21} developed a proof-of-concept JRC platform using low-resolution A/Ds at mmWave frequencies, which reduced the energy consumption as well as the hardware cost. {Nevertheless, the quantized design of A/Ds inevitably leads to performance degradation of both radar and communication subsystems, making it undesirable especially at low-THz frequencies with low SNR at the receiver.} {It is evident that, the existing literature has not completely addressed all the above-mentioned technical barriers encountered in the mmWave/low-THz scale, e.g., severe issues of path loss, Doppler shifts and implementation cost.} To this end, this paper extensively investigates the waveform design issue for JRC systems with ultra-broad bandwidth at mmWave and low-THz frequencies, where the main contributions can be summarized as follows: \\
1) To alleviate the harsh requirement on the A/D sampling rate in JRC systems with ultra-broad bandwidth, a multi-subband quasi-perfect (MS-QP) sequence is proposed for radar sensing, {which is obtained from the superposition of multiple perfect subsequences (e.g., ZC sequence, generalized Frank sequence and Milewski sequence \cite{Zhang_isit_19})} on different subbands. Here the perfect sequence is referred to as the sequence whose periodic auto-correlation value always equals zero for any non-zero shift\cite{Zhang_isit_19}. By such arrangement, it is capable of maintaining ultra-broad bandwidth with good auto-correlation property, whilst only low-rate A/Ds are required for signal recovery at the radar receiver, leading to reduced hardware cost. {For clarity, ZC subsequences \cite{AlSharif_conf_17} are mainly considered for MS-QP sequence generation in this paper as a special case.}\\
2) The parameters of the proposed MS-QP sequence {based on ZC subsequences} are specially designed against strong Doppler shift in the mmWave and THz scale. More specifically, by considering each ZC subsequence on different subbands for radar sensing respectively, its root index is optimized to gather high sidelobes induced by Doppler shift closely around the mainlobe on the radar range profile\footnote{Range profile is defined as the cross-correlation results between the echo signals and the transmitted sensing sequence. Correspondingly, range sidelobes are referred to as the sidelobes besides the main peak on the range profile.}, which also leads to well concentration of dominant range sidelobes to the main peaks of the proposed MS-QP sequence. Afterwards, a target detection criterion called Target Exclusion nearby the Main Peak (TEMP) is invoked assuming absence of other target nearby the main peaks, thus mitigating the false alarms caused by Doppler shift. \\
3) An data-embedded MS-QP (DE-MS-QP) waveform is proposed based on the MS-QP sequence, which enables simultaneous high-resolution sensing and high-rate communication without mutual interference. More specifically, the MS-QP sequence is repetitively transmitted to generate null frequency points on each subband available for data transmission. Then the data symbols are inserted into these frequency points through time-domain replication and phase rotation procedures, yielding the proposed DE-MS-QP waveform. The resource allocation strategy of sensing and communication can be flexibly adjusted to reach a desirable performance trade-off between the two functions, by simply changing the waveform parameter. \\
4) Simulation results are provided to validate the feasibility of the proposed parameter design against Doppler shift, and that the proposed waveforms are capable of achieving ultra-high-resolution ranging and velocity estimation under extremely noisy environment, with cheaper front-end devices than classical wide-band ZC sequences and linear frequency modulated (LFM) signals. Moreover, the resource allocation strategy between the dual functions of the proposed DE-MS-QP waveform can be optimized numerically to enable simultaneous mm-level radar sensing and data transmission with marginal degradation of the communication performance.

Before we proceed, we point out that this paper places an emphasis on the aspect of radar sensing in the proposed JRC waveform design.

\subsection{Structure and Notations}
The remainder of the paper is organized as follows. Section \ref{s2} illustrates the JRC system model. Section \ref{s3} presents the proposed waveform design, which is followed by the corresponding receiving techniques in Section \ref{s4}. Numerical performance evaluation is then provided in Section \ref{s5}, and Section \ref{s6} draws the conclusion.

\emph{Notation}: $(\cdot)^*$, $\|\cdot\|$ and $\left |\cdot  \right |$ denote the conjugate operator, the 2-norm, and the modulus of a set, respectively. $\langle \cdot \rangle_N$ stands for the modulo-$N$ operator.

\section{System Model}\label{s2}
Figure \ref{fig1} illustrates the architecture of the considered JRC system, which characterizes typical JRC applications such as holographic conferencing with gesture detection and downlink data transfer, and vehicle-to-vehicle sensing and communication for automated driving. Both of the radar and communication subsystems share the same hardware platform with co-located transceivers. For sensing/communication purposes, $\mathbf{x}=\left [x[0],x[1],\cdots,x[N-1]\right ]$ are subsequently fed into a digital-to-analog converter (D/A), a mixer and a high power amplifier (HPA) to generate analog signals for transmission. To overcome the severe path loss at mmWave and THz frequencies, high-gain directional antennas (e.g., horn antennas, parabolic antennas, and lens antennas) with extremely narrow beams are usually employed at the transceiver \cite{Alouini_survey_2020,Rappaport_survey_19}, where the multi-path effects could be limited. Therefore, a line-of-sight (LoS) flat-fading channel is assumed between the JRC platform and the targets \cite{Liu_tcom_20,Han_tsp_16} with non-dominant clutters on the path treated as additive noise \cite{Kumari_tvt_18}. Except for the directional antennas, the utilization of large antenna arrays with beamforming techniques could also compensate for the path loss effectively. When each antenna element transmits the same signal for diversity gain, the overall array with beamforming is actually equivalent to a single directional antenna. Hence, we consider a single-input single-output (SISO) scenario for simplicity.

After arriving at the targets, on one hand, the conveyed data can be extracted from the received waveform by the mobile devices. On the other hand, $\mathbf{x}$ is also bounced back to the JRC platform for receive (RX) signal processing including down-conversion and A/D operations. The resultant baseband echoes, denoted as $\mathbf{y}=\left [y[0],y[1],\cdots,y[N-1]\right ]$, is then utilized for target sensing via time-domain correlation-based methods\footnote{Aside from radar applications, the RX branch of the JRC platform is also capable of demodulating the uplink signals from the the users/targets as shown in Fig. \ref{fig1}, where sensing and uplink communication are assumed to follow TDD working mode.}. Aside from the round-trip delay as well as the Doppler shift effects, $\mathbf{y}$ inevitably suffers from hardware impairment including in-phase/quadrature (I/Q) imbalance, phase noise and HPA nonlinearity induced by the mmWave/THz front ends \cite{Xiao_jsac_17,Boulogeorgos_access_19}, due to the difficulty in the fabrication of ultra-high-frequency devices. For simplicity, we assume that I/Q imbalance and nonlinearity of HPA at the transmit (TX) side have been removed with pre-compensation techniques \cite{Chung_tmtt_18,Ramadan_access_18}. {Then, the incoming signal before the mixer at the radar receiver, denoted as $y_0[n]$, can be formulated as
\begin{equation}\label{eq1}
	y_0[n]\approx \sum_{i=1}^{I}\left(h_ix[n-\tau_i]e^{\textsf{j}(2\pi n{v}_i+\theta_{n-\tau_i})}\right)+w[n],
\end{equation}
where $I$ is the number of available targets, and $h_i$ denotes the round-trip path gain of $i$th target, equal to the multiplication of antenna gains, free-space propagation loss, molecular absorption loss as well as the reflection loss \cite{Boulogeorgos_access_19}. $w[n]$ is a complex Gaussian variable of $\mathcal{CN}(0,\sigma^2)$, defined as the aggregate noise component of thermal noise and clutters. Besides, $\tau_i=\left \lfloor t_i/T_s \right \rfloor$ is defined as the integer delay with $t_i$ and $T_s$ representing the round-trip delay of the $i$th target and the sampling period, respectively. Note that the fractional part of the time delay due to finite sampling rate is omitted in (\ref{eq1}) for brevity. In addition, ${v}_i$ denotes the normalized Doppler shift, formulated as
\begin{equation}
	{v}_i=\frac{2u_if_c}{c_0}T_s,
\end{equation}
where $u_i$, $f_c$ and $c_0$ are defined as the relative speed of the $i$th target, the center frequency of the JRC system and the speed of light, respectively. $\theta_n$ for $n=0,1,\cdots,N-1$ is the phase noise term of the local oscillator (LO) shared by the co-located transceiver, following a random-walk model (Wiener process) illustrated as \cite{Colavolpe_jsac_05}
\begin{equation}
	\theta_n=\theta_{n-1}+\Delta\theta_n,\:\:n=1,2,\cdots,N-1,
\end{equation}
where $\Delta\theta_n\sim \mathcal{N}(0,\sigma_\Delta^2)$ denotes the Gaussian random variation of phase noise, and $\theta_0$ is often set to be uniformly distributed in $[0,2\pi)$. Afterwards, $y_0[n]$ is fed into the mixer for down-conversion, yielding
\begin{equation}\label{eq2}
	y[n]\approx\mu _ry_0[n]e^{-\textsf{j}\theta_n}+\nu_ry^*_0[n]e^{\textsf{j}\theta_n},
\end{equation}
given that
\begin{equation}\label{eq3}
	\left\{\begin{matrix}
		\mu_r=\cos\phi_r+j\epsilon_r\sin\phi_r;\\
		v_r=\epsilon_r\cos\phi_r-j\sin\phi_r,
	\end{matrix}\right.
\end{equation}
where $\phi_r$ and $\epsilon_r$ denote the phase and amplitude imbalances of the RX I/Q branches, respectively \cite{Jiang_cl_13}. By substituting (\ref{eq1}) into (\ref{eq2}), the echo signal for radar processing can be formulated as
\begin{equation}
\begin{aligned}
y[n]\approx&\mu _r\left(\sum_{i=1}^{I}\left(h_ix[n-\tau_i]e^{\textsf{j}(2\pi n{v}_i+\theta_{n-\tau_i})}\right)+w[n]\right)e^{-\textsf{j}\theta_n}+\\&\nu_r\left(\sum_{i=1}^{I}\left(h^*_ix^*[n-\tau_i]e^{-\textsf{j}(2\pi n{v}_i+\theta_{n-\tau_i})}\right)+w^*[n]\right)e^{\textsf{j}\theta_n}\\&\hspace{-4mm}=\sum_{i=1}^{I}\Big(h_i\mu _rx[n-\tau_i]e^{\textsf{j}(2\pi n{v}_i+\theta_{n-\tau_i}-\theta_n)}+\\&h_i^*\nu _rx^*[n-\tau_i]e^{-\textsf{j}(2\pi n{v}_i+\theta_{n-\tau_i}-\theta_n)}\Big)+\tilde{w}[n],
\end{aligned}
\label{eq4}
\end{equation}
where $\tilde{w}[n]$ denotes the equivalent noise term involving the impacts of hardware imperfections.}

As is observed in (\ref{eq4}), the received echoes are distorted by hardware imperfections. However, unlike the wireless communication subsystem at mmWave and low-THz frequencies, which is vulnerable to hardware impairment, I/Q imbalance and LO phase noise are less dominant in radar sensing. Explicitly, the echo signals are composed of the received sensing sequence as well as its image component induced by I/Q imbalance. The power of the former is usually much stronger than the image counterpart \cite{Kallfass_jimtw_15,Kallfass_te_15,Grzyb_conf_18}, whose impacts are marginal on the correlation results. Besides, since a single LO is shared by the co-located radar transceiver, the TX and RX phase noise terms in (\ref{eq1}) could cancel each other out approximately if the round-trip delay is sufficiently small. Such condition is readily satisfied for JRC systems at mmWave and low-THz frequencies, which mainly serve short-range target/users due to the severe path loss. Therefore, hardware imperfections will not be an emphasis in our JRC waveform design.
\begin{figure*}[t!]
	\begin{center}
		\includegraphics[width=0.9\linewidth, keepaspectratio]{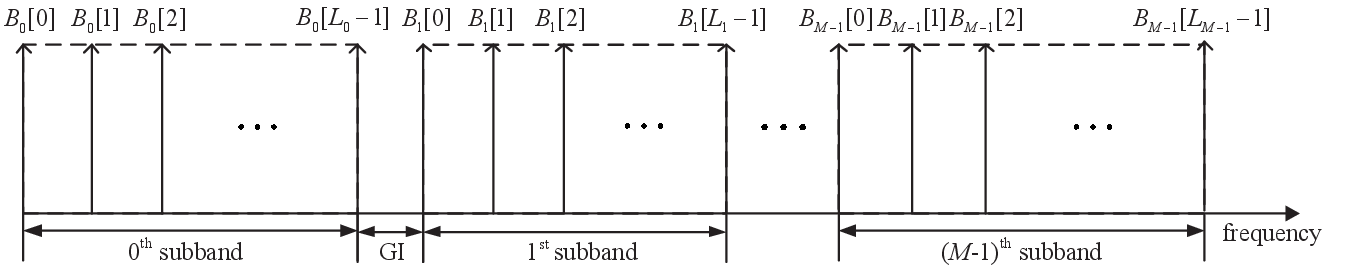}
	\end{center}
	\caption{Frequency-domain representations of the proposed MS-QP sequence.}
	\label{fig2}

\end{figure*}
\section{Proposed JRC Waveform Design in mmWave and Low-THz Bands}\label{s3}
\subsection{MS-QP Sequence for Radar Sensing}\label{s3.1}
Since ultra-broad bandwidth is employed in the mmWave/low-THz JRC system to achieve both high-resolution ranging and data transmission over $100$ Gbps, A/Ds with sampling rate over tens of GHz are required in principle at the receiver, leading to undesirable hardware cost. To this end, a novel MS-QP sequence is constructed for radar applications using multiple perfect sequences (e.g., ZC sequences \cite{Frank_tit_62,Chu_tit_72}) on different frequency subbands, which could support ultra-high-resolution ranging with the only requirement of low-rate A/Ds. Specifically, a set of $M$ ZC sequences\footnote{For clarity, ZC sequences are employed for MS-QP sequence generation without loss of generality. The construction method is also applicable to the rest of the perfect sequence family, e.g., generalized Frank sequence, Milewski sequence, etc. \cite{Zhang_isit_19}.}, $\mathbf{b}_m=\left[b_m[0],b_m[1],\cdots,b_m[L_m-1]\right]$ for $m=0,1,\cdots,M-1$, is generated for MS-QP sequence construction, each formulated as
\begin{equation}\label{eq5}
b_m[l]=\exp\left ( -\textsf{j}\pi\frac{p_ml(l+1)}{L_m} \right ),
\end{equation}
with ideal auto-correlation property shown as
\begin{equation}\label{eq6}
r_{b_mb_m}[n]=\sum_{l=0}^{L_m-1}b_m[l]b_m^*[l-\tau+n]=\left\{\begin{matrix}
L_m, & n=\tau;\\
0, & n\neq \tau,
\end{matrix}\right.
\end{equation}
where {$l$ and $n$ denote the indices of the ZC sequences and correlation results, respectively.} $r_{b_mb_m}[n]$ denotes the auto-correlation value, and $\tau$ represents the round-trip delay. Besides, $L_m$ and $p_m$ are defined as the length of $\mathbf{b}_m$ (odd number) and the root index, respectively, following $\gcd(p_m,L_m)=1$. Afterwards, $\mathbf{b}_m$ for $m=0,1,\cdots,M-1$ are converted to the frequency domain using $L_m$-point discrete Fourier transform (DFT), yielding $B_m[k]$ with {the index $k=0,1,\cdots,L_m-1$.} The frequency components of $\mathbf{b}_m$ are then moved to the $m$-th frequency subband for $m=0,1,\cdots,M-1$, respectively, where a guard interval (GI) of length $L_G$ is inserted between any two neighboring subbands\footnote{As stated in Section \ref{s4}, $M$ band-pass filters are employed to obtain the frequency components of each subband at the radar receiver. To avoid out-of-band interference on the filtered results of each subband, $L_G$ should be designed to guarantee that, the 3-dB bandwidth of the band-pass filter employed for each frequency subband is non-overlapped with its neighbouring subbands.}, as illustrated in Fig. \ref{fig2}. Finally, an inverse DFT (IDFT) with size of $N=(\sum_{m=0}^{M-1}L_m)+ML_G$ is imposed on the concatenated spectrum of the $M$ subbands, yielding the equivalent baseband representation of the proposed MS-QP sequence, derived as
\begin{equation}\label{eq7}
{\begin{aligned}
x[n]&\overset{\triangle }{=}\frac{1}{\sqrt{N}}\sum_{m=0}^{M-1}e^{\frac{\textsf{j}2\pi f_mn}{N}}\sum_{k=0}^{L_m-1}B_m[k]e^{\frac{\textsf{j}2\pi}{N}kn}\\&=\frac{1}{\sqrt{N}}\sum_{m=0}^{M-1}e^{\frac{\textsf{j}2\pi f_mn}{N}}\frac{1}{\sqrt{L_m}}\sum_{l=0}^{L_m-1}b_m[l]\times\\&\frac{\sin\left ( L_m\pi(\frac{n}{N}-\frac{l}{L_m}) \right )}{\sin\left ( \pi(\frac{n}{N}-\frac{l}{L_m}) \right )}\times e^{\textsf{j}\pi(L_m-1)(\frac{n}{N}-\frac{l}{L_m})},
\end{aligned}}
\end{equation}
where the frequency shift $f_m$ can be calculated by
\begin{equation}\label{eq8}
f_m=\left\{\begin{matrix}
(\sum_{i=0}^{m-1}L_{i})+mL_G, & 1\leq m\leq M-1;\\
0, & m=0.
\end{matrix}\right.
\end{equation}
{Since the proposed MS-QP sequence is constructed by concatenating multiple subbands separated by frequency-domain guard interval, it could utilize ultra-broad bandwidth for high-resolution ranging with the only requirement of $M$ cost-efficient low-rate A/Ds for signal sampling on each subband independently, instead of full-band A/D with the sampling rate over tens of GHz\footnote{The proposed multi-subband structure can also be combined with the classical single-carrier communication waveform, realizing straightforward communication and sensing convergence with low expenses, which however could not ensure good auto-correlation property, and the Doppler-resilient design like Section \ref{s3.2} is also infeasible due to the randomness of communication signals.}. Thus the hardware expenses can be eliminated.} On the other hand, to investigate the auto-correlation property of the proposed MS-QP sequence, a useful lemma already demonstrated in \cite{Popovic_tit_18} is introduced as below:

{\bf Lemma 1}: \cite{Popovic_tit_18} \emph{An arbitrary sequence whose DFT coefficients are of constant magnitude, has ideal auto-correlation property, and vice versa.}

It is readily proved that the magnitudes of DFT coefficients of any ZC sequence are constant. Hence, the spectrum envelope of the proposed MS-QP sequence, constituted by $B_m[k]$ for $k=0,1,\cdots,L_m-1$ and $m=0,1,\cdots,M-1$, is nearly constant if $L_G$ is sufficiently small. This intuitively implies that the proposed MS-QP sequence could achieve quasi-perfect auto-correlation property according to the lemma, making it {more qualified} for radar sensing applications.

\begin{figure*}
	\centering
	\subfigure[Sequence generator based on digital circuits.]{
		\label{fig3a} 
		\includegraphics[width=0.4\linewidth]{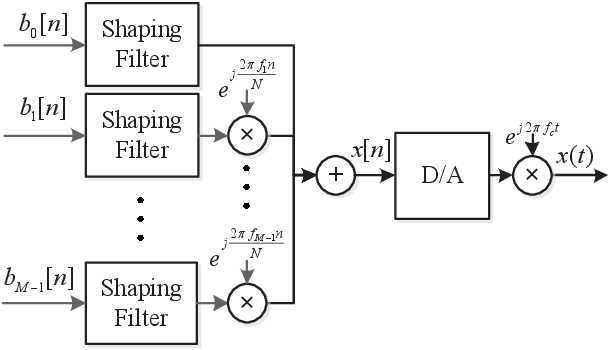}}
	\subfigure[Sequence generator based on analog circuits. ]{
		\label{fig3b} 
		\includegraphics[width=0.37\linewidth]{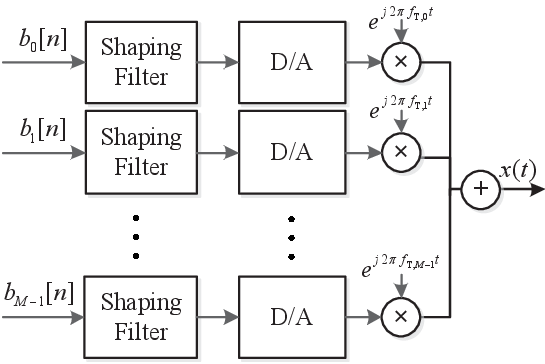}}
	\caption{Transmitter diagram for MS-QP sequence generation.}
	\label{fig3} 
	
\end{figure*}

Despite the aforementioned merits of the proposed MS-QP sequence, the time-domain constant-amplitude property is unfortunately not inherited from its ZC sequence components, which may cause high PAPR problem. This could be detrimental especially for high-rate mmWave/THz systems with HPAs \cite{Bjornson_tcom_13}. In order to address this issue, whilst retaining the quasi-perfect auto-correlation property, we recall that the auto-correlation property of a sequence is mainly dependent on its amplitude-frequency characteristic \cite{Zeng_tvt_20}, which enlightens us to alter the phase-frequency characteristic of the proposed MS-QP sequence for PAPR reduction. More specifically, a phase rotation of $\hat{\phi}_m$ chosen from an $L_\phi$-element alphabet denoted as $\Phi =\{\phi_0,\phi_1,\cdots,\phi_{L_\phi-1}\}$, is imposed on the $m$th frequency subband for $m=0,1,\cdots,M-1$. Hence, the proposed MS-QP sequence $x[n]$ for $n=0,1,\cdots,N-1$, originally given in (\ref{eq7}), can be finally modified as
\begin{equation}\label{eq9}
\begin{aligned}
x[n]&\overset{\triangle }{=}\frac{1}{\sqrt{N}}\sum_{m=0}^{M-1}e^{\textsf{j}\left(\frac{2\pi f_mn}{N}+\hat{\phi}_m\right)}\frac{1}{\sqrt{L_m}}\times\\&\sum_{l=0}^{L_m-1}b_m[l]\frac{\sin\left ( L_m\pi(\frac{n}{N}-\frac{l}{L_m}) \right )}{\sin\left ( \pi(\frac{n}{N}-\frac{l}{L_m}) \right )} e^{\textsf{j}\pi(L_m-1)(\frac{n}{N}-\frac{l}{L_m})},
\end{aligned}
\end{equation}
where we have
\begin{equation}\label{eq10}
\left [ \hat{\phi}_0,\hat{\phi}_1,\cdots,\hat{\phi}_{M-1} \right ]=\arg\underset{\hat{\phi}_i\in\Phi}{\min}\frac{\max(\left \| {x}[n] \right \|^2)}{E(\left \| {x}[n] \right \|^2)}.
\end{equation}

Finally, for practical implementations to generate MS-QP sequences, one straightforward solution is shown in Fig. \ref{fig3a}, where $\mathbf{b}_m$ for $m=0,1,\cdots,M-1$ with sampling period $T_m$ {performs interpolation and pulse shaping through the shaping filter,} then moved to different subbands in the low-frequency scale, and finally superposed together to generate the MS-QP sequence $\mathbf{x}$ with sampling period $T_s$ through digital circuits. Here we have $T_m=NT_s/L_m$ for $m=0,1,\cdots,M-1$. Such method however requires an ultra-high-rate D/A to generate analog radar sensing signals $x(t)$. Alternatively, as illustrated in Fig. \ref{fig3b}, $\mathbf{b}_m$ can be directly converted into analog signals through low-rate D/As, and then moved to its corresponding subband with center frequency ${f}_{\text{T},m}$ in the mmWave/THz scale, respectively. Finally, different branches are superposed to generate $x(t)$.

{\bf Remark 1}: Aside from sensing, the transmitter architecture of the proposed MS-QP sequence is also capable of supporting 5th-generation (5G) and beyond wireless communications, e.g., the multi-service subband filtered multi-carrier (MS-SFMC) scheme \cite{Zhang_twc_17}. In MS-SFMC systems, the overall bandwidth is split into multiple subbands separated by GIs utilized for different data services, where the bandwidth allocation for different subbands can be flexible, and data on each subband is extracted for a particular service/user using band-pass filtering. This fits quite well with the MS-QP sequence transmitter in Fig. \ref{fig3}, making the proposed MS-QP sequence readily available for JRC applications in a TDD manner.

\subsection{Parameter Design for MS-QP Sequences against Strong Doppler Shift}\label{s3.2}
Doppler shift tends to be stronger at mmWave/THz frequencies, which induces dominant sidelobes on the range profile of radar sensing, leading to additional false alarms at the radar receiver. Hence, the parameters of the proposed MS-QP sequence are required to be specially designed in order to enhance its robustness against Doppler shift. It can be observed in (\ref{eq9}) that the MS-QP sequence is the superposition of $M$ subsequences $\tilde{\mathbf{b}}_m=\left[\tilde{b}_m[0],\tilde{b}_m[1],\cdots,\tilde{b}_m[N-1]\right]$ for $m=0,1,\cdots,M-1$ on non-overlapped subbands, which are expressed by
\begin{equation}\label{eq1_remark1}
\begin{aligned}
\tilde{b}_m[n]&{=}\frac{1}{\sqrt{NL_m}}e^{\textsf{j}\left(\frac{2\pi f_mn}{N}+\hat{\phi}_m\right)}\sum_{l=0}^{L_m-1}b_m[l]\\&\times\frac{\sin\left ( L_m\pi(\frac{n}{N}-\frac{l}{L_m}) \right )}{\sin\left ( \pi(\frac{n}{N}-\frac{l}{L_m}) \right )} e^{\textsf{j}\pi(L_m-1)(\frac{n}{N}-\frac{l}{L_m})}.
\end{aligned}
\end{equation}
It is indicated that $\tilde{\mathbf{b}}_m$ for $m=0,1,\cdots,M-1$ are approximately orthogonal, i.e., having zero cyclic cross-correlations, since their used spectrums are non-overlapped. The orthogonality still holds even under Doppler shift effects by setting guard interval of appropriate length between neighbouring subbands. This indicates that the range sidelobes of the MS-QP sequence induced by Doppler shift is mainly determined by the cross-correlation results between $\tilde{\mathbf{b}}_m$ and its corresponding echo signal for $m=0,1,\cdots,M-1$. Furthermore, it is seen from (\ref{eq1_remark1}) that $\tilde{\mathbf{b}}_m$ is actually an interpolated and frequency-shifted ZC sequence, which is expected to share similar properties with $\mathbf{b}_m$, including the distribution of sidelobes induced by Doppler shift on the range profile. Therefore, to overcome the Doppler shift effects on radar sensing with the MS-QP sequence, a heuristic strategy is to mitigate its impacts on the range profile of each ZC subsequence $\mathbf{b}_m$ ($m=0,1,\cdots,M-1$) separately by special design of the corresponding root index $p_m$.

Assume that an arbitrary ZC sequence $\mathbf{b}=\left[b[0],b[1],\cdots,b[N_{zc}-1]\right]$ is invoked for radar sensing, which is formulated as
\begin{equation}
b[n]=\exp\left ( -\textsf{j}\pi\frac{pn(n+1)}{N_\text{zc}} \right ),
\end{equation}
where $N_\text{zc}$ and $p$ denote the length (odd) and root index of the ZC sequence, respectively, satisfying $0<p<N_\text{zc}$ and $\gcd(p,N_\text{zc})=1$. It is indicated that, the ideal auto-correlation property of $\mathbf{b}$ can be impaired by the Doppler shift effects. Such impact is more pronounced in mmWave/THz systems than the lower-frequency counterparts, due to high operating frequency at the order of $100$ GHz. Explicitly, by considering the single-target scenario for brevity, the magnitude of cross-correlation between the transmitted ZC sequence and the echoes with round-trip delay $\tau$ and normalized Doppler shift $v$, can be derived as ($n=0,1,\cdots,N_{\text{zc}}-1$)
\begin{equation}
\begin{aligned}
\left \| r_{bb}[n] \right \|&=\left \| \sum_{k=0}^{N_{\text{zc}}-1}b[k-\tau]e^{\textsf{j}2\pi {v}k}b^*[k-n] \right \|
\\&=\left \| \frac{\sin\left(\pi\left \langle p(n-\tau)-{v}N_\text{zc} \right \rangle_{N_{\text{zc}}}\right)}{\sin\left( \frac{\pi}{N_\text{zc}}\left \langle p(n-\tau)-{v}N_\text{zc} \right \rangle_{N_{\text{zc}}}\right )} \right \|,
\end{aligned}
\end{equation}
which can be seen as a function of $\left \langle p(n-\tau)-{v}N_\text{zc} \right \rangle_{N_{\text{zc}}}$\footnote{Note that the remainder obtained by the modulo operator $\left \langle \cdot \right \rangle_{N_{\text{zc}}}$ could be a fractional number.}, as illustrated in Fig. \ref{fig4}. When the Doppler shift is marginal, i.e., ${v}\approx 0$, it is observed that $\left \langle p(n-\tau)-{v}N_\text{zc} \right \rangle_{N_{\text{zc}}}$ becomes integer for arbitrary value of $n$, and thus $\left \| r_{bb}[n] \right \|$ is equal to zero except for $n=\tau$ corresponding to the main peak on the range profile. On the other hand, under non-negligible Doppler shift, the values of $\left \langle p(n-\tau)-{v}N_\text{zc} \right \rangle_{N_{\text{zc}}}$ for $n=0,1,2,\cdots,N_{\text{zc}}-1$ deviate from the integer points, leading to non-zero range sidelobes. By assuming that ${v}N_\text{zc}< 1$\footnote{When ${v}N_\text{zc}\geq 1$, the strong Doppler shift could even cause translation of the main peak on the range profile, aside from the generation of range sidelobes, leading to inevitable error floor for target ranging. In this paper, the Doppler shift is partly mitigated by periodical transmission of shorter codes in each coherent processing interval (CPI) instead of directly using long codes \cite{Zeng_tvt_20}, where ${v}N_\text{zc}< 1$ is satisfied.}, it is seen that these sidelobes are higher at the time instants $n$ corresponding to smaller $\left\|\left \langle p(n-\tau)\right \rangle_{N_{\text{zc}}}\right\|$.

On the other hand, since $\gcd(p,N_\text{zc})=1$, $\mathbf{R}=\left[\left \langle -\tau p\right \rangle_{N_{\text{zc}}},\left \langle (1-\tau)p\right \rangle_{N_{\text{zc}}},\cdots,\left \langle (N_\text{zc}-1-\tau)p\right \rangle_{N_{\text{zc}}}\right]$ constitutes a complete system of residues with respect to $N_\text{zc}$, constrained to be within $\left [ -\frac{N_{\text{zc}}-1}{2},\frac{N_{\text{zc}}-1}{2} \right ]$ without loss of generality. In other words, $\mathbf{R}$ can be regarded as a permutation of all the $N_{\text{zc}}$ integers in $\left [ -\frac{N_{\text{zc}}-1}{2},\frac{N_{\text{zc}}-1}{2} \right ]$, where each element corresponds to a unique sidelobe on the range profile as shown in Fig. \ref{fig4}. By changing the root index, the permutation can be modified, and the position of sidelobes on the range profile can then be altered accordingly. Based on the aforementioned findings, a novel root index design for ZC sequences against Doppler shift is proposed as follows, aiming at concentrating the dominant sidelobes induced by Doppler shift closely around the main peak on the range profile: \emph{For any odd-length ZC sequence $b[n]$ ($n=0,1,\cdots,N_\text{zc}-1$), the root index could be set as $p\in\{1,\frac{N_{\text{zc}}-1}{2},\frac{N_{\text{zc}}+1}{2},N_\text{zc}-1\}$ if satisfying $\gcd(p,N_\text{zc})=1$.}

\begin{figure}[t!]
	\begin{center}
		\includegraphics[width=1\linewidth, keepaspectratio]{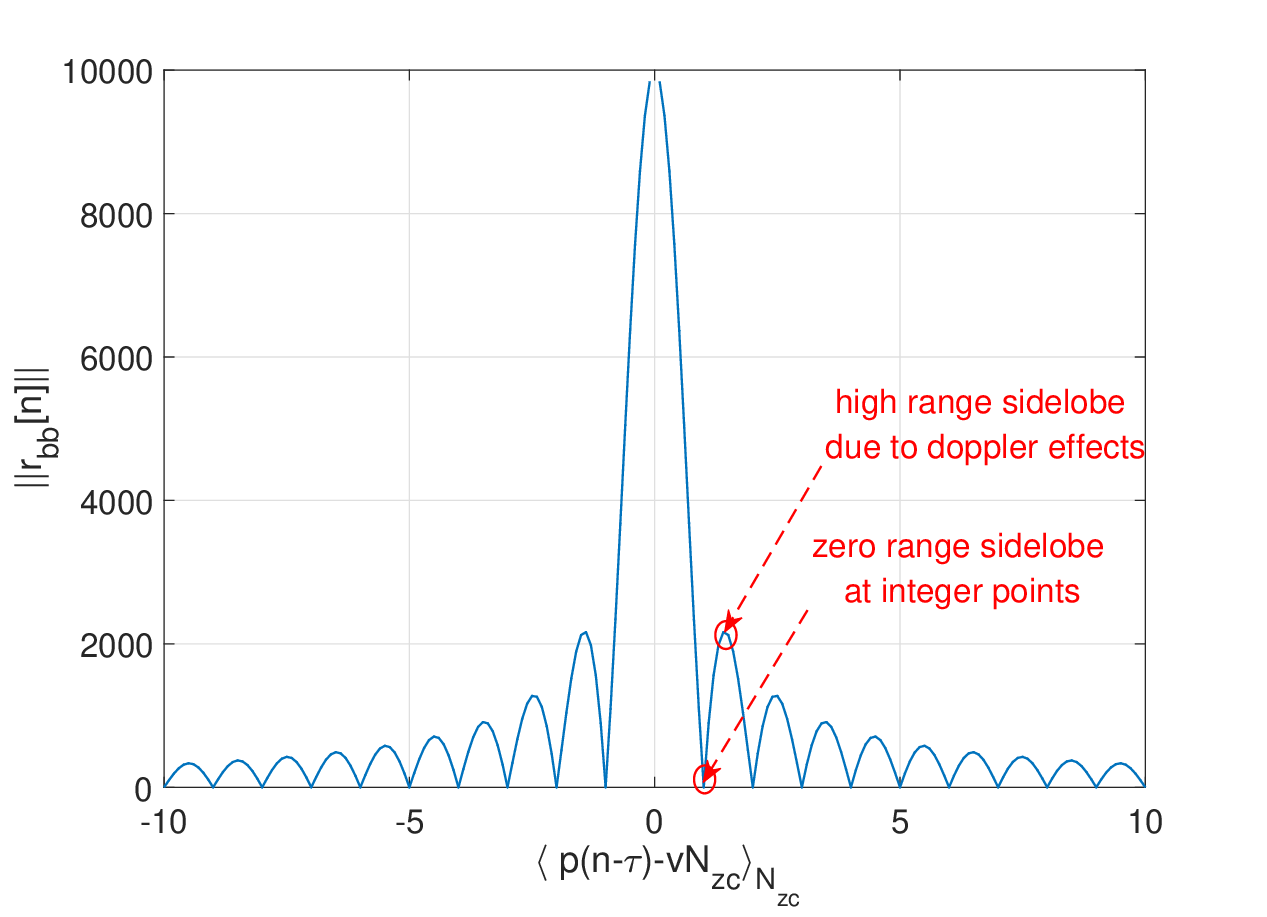}
	\end{center}
	\caption{The graph of $\left \| r_{bb}[n] \right \|$ as a function of $\left \langle p(n-\tau)-{v}N_\text{zc} \right \rangle_{N_{\text{zc}}}$, where $N_\text{zc}=10007$.}
	\label{fig4}
\end{figure}

To validate the feasibility of the proposed root index design, we consider the cases of $p=1$ and $p=\frac{N_{\text{zc}}-1}{2}$, where the elements of $\mathbf{R}$ can be respectively calculated as
\begin{equation}\label{eq13.5}
\left \langle n-\tau\right \rangle_{N_\text{zc}}= n-\tau,
\end{equation}
\begin{equation}\label{eq14}
\left \langle \frac{N_{\text{zc}}-1}{2}(n-\tau) \right \rangle_{N_\text{zc}}= \left\{\begin{matrix}
-k, & n-\tau=2k; \\
\frac{(N_{\text{zc}}+1)}{2}-k,  & n-\tau=2k-1.
\end{matrix}\right.
\end{equation}
It is seen from (\ref{eq13.5}) and (\ref{eq14}) that the time instants of $n$ corresponding to smaller $\left \langle  p(n-\tau)\right \rangle_{N_\text{zc}}$ are distributed at both sides of $n=\tau$ closely, making the dominant sidelobes well concentrated around the main peak on the range profile. On one hand, such property is apparent for $p=1$; On the other hand, Table \ref{t1} with $N_\text{zc}=10007$ and $p=5003$ exemplifies the case of $p=\frac{N_{\text{zc}}-1}{2}$, where the distances from the main peak to the highest and second-highest sidelobes are $2$ and $4$, respectively. Fig. \ref{fig5} evaluates the effectiveness of our proposed root index design, given that $N_\text{zc}=10007$, and $\tau=1000$. Compared with the case of $p=3$ in Fig. \ref{fig5a}, where non-negligible sidelobes are witnessed away from the main peak, dominant sidelobes are closely distributed around the main peak for the proposed parameter design, as shown in Figs. \ref{fig5b} and \ref{fig5c}. Similar results can be obtained for $p=N_\text{zc}-1$ and $p=\frac{N_{\text{zc}}+1}{2}$, which are omitted here for brevity.

\begin{figure*}[t!]
	\centering
	\subfigure[$p=3$.]{
		\label{fig5a} 
		\includegraphics[width=0.31\linewidth]{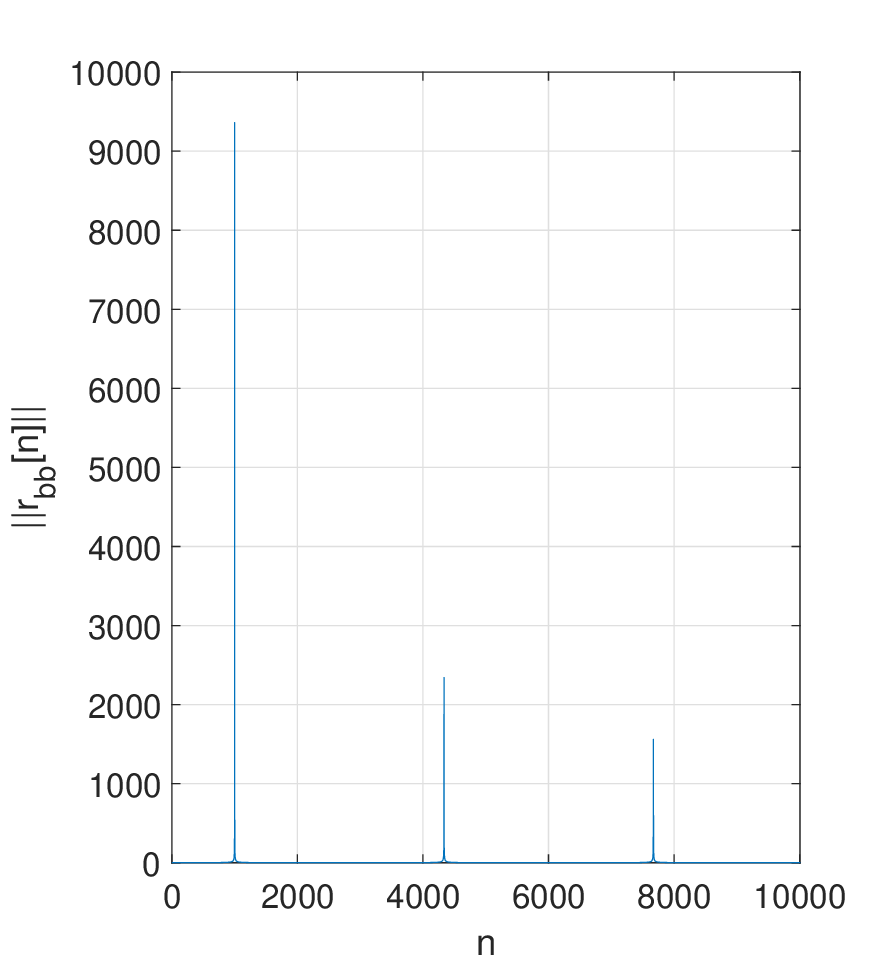}}
	\subfigure[$p=1$ (proposed).]{
		\label{fig5b} 
		\includegraphics[width=0.31\linewidth]{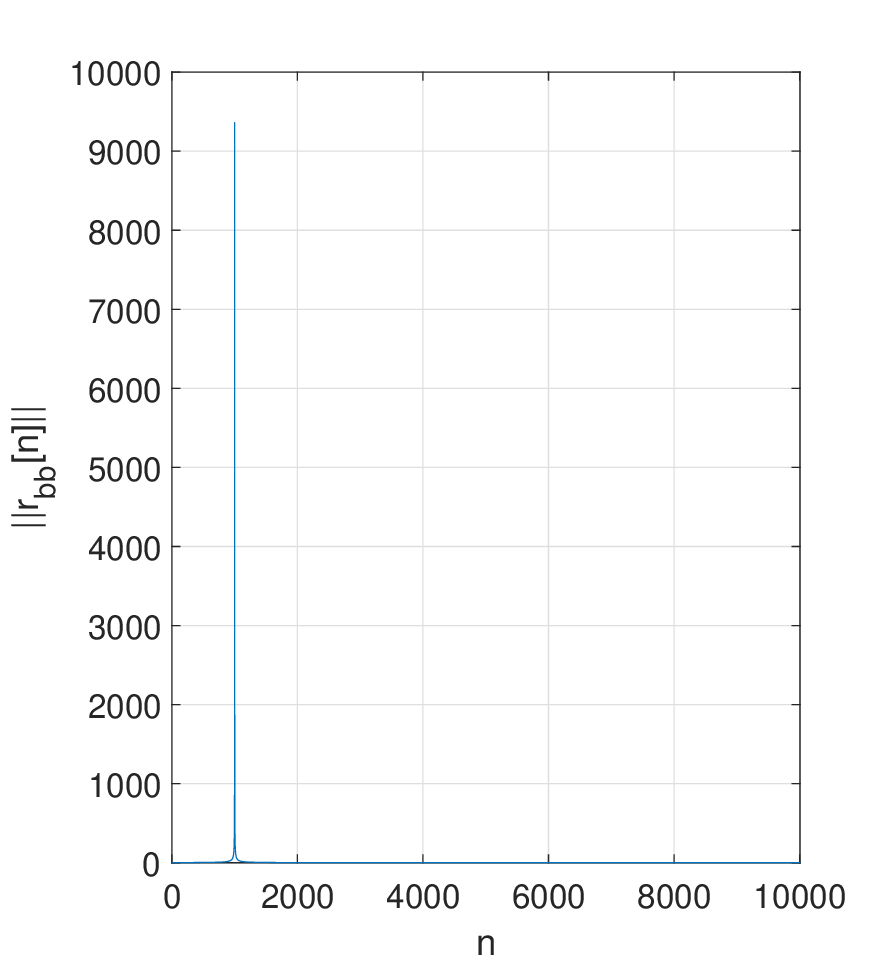}}
	\subfigure[$p=5003$ (proposed).]{
		\label{fig5c} 
		\includegraphics[width=0.31\linewidth]{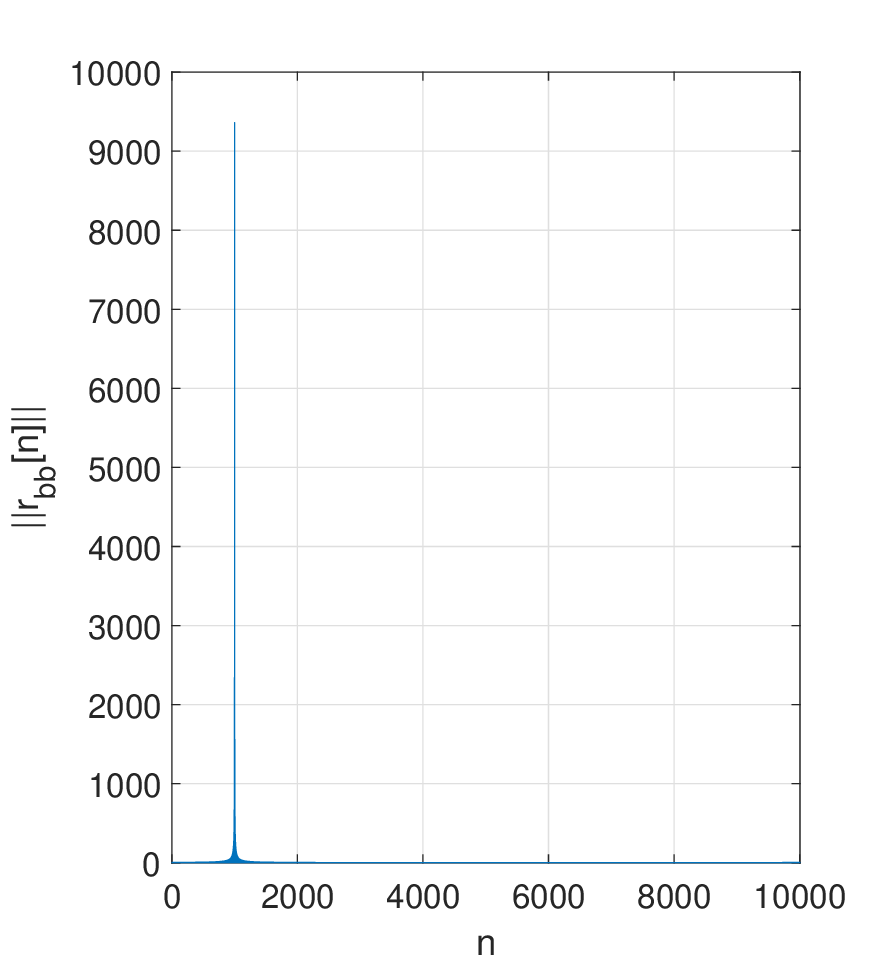}}
	\caption{Range profile of the ZC sequences with and without proposed root index design, where $N_\text{zc}=10007$, and $\tau=1000$.}
	\label{fig5} 
\end{figure*}

\begin{table*}[t!]
\renewcommand{\arraystretch}{1.1}
\caption{Distribution of the Sidelobes Induced by Doppler Shift in the Range Profile of the ZC Sequence with $N_\text{zc}=10007$ and $p=5003$.}
\centering
\begin{tabular}{|c | c | c | c | c | c | c | c | c | c | c |c|}
\hline
 $\left \langle  n-\tau \right \rangle_{N_\text{zc}}$ &$\cdots$ & $-4$ & $-3$ & $-2$ & $-1$ & $0$ & $1$ & $2$ & $3$ & $4$ &$\cdots$\\\hline
 $\left\|\left \langle  p(n-\tau)\right \rangle_{N_\text{zc}}\right\|$  &$\cdots$  &{2}& 5002 & {1} & 5003 & {0} & 5003 & {1} & 5002 & {2}&$\cdots$ \\\hline
 Sidelobe Level & $\cdots$ & {High} & Low & {High} & Low& {Peak} & Low & {High} & Low & {High} & $\cdots$\\\hline
\end{tabular}
\label{t1}
\end{table*}

As discussed above, when applying the MS-QP sequence to radar sensing, its sidelobe distribution on the range profile is determined by those of its ZC subsequences $\mathbf{b}_m$ for $m=0,1,\cdots,M-1$. Hence, the range sidelobes caused by the Doppler shift can be well concentrated around the main peaks on the range profile of the proposed MS-QP sequence, by setting $p_m\in\{1,\frac{L_m-1}{2},\frac{L_m+1}{2},L_m-1\}$ for $m=0,1,\cdots,M-1$. Afterwards, the target detection scheme can be specially designed to eliminate the possible false alarms caused by the strong Doppler shift, which will be discussed into detail in Section \ref{s4}.


\subsection{Integrated DE-MS-QP Waveform for JRC Applications}\label{s3.3} 
As stated in {\bf Remark 1}, the proposed MS-QP sequence is also compatible with wireless communication function under TDD working mode, which however cannot support concurrent radar sensing and data transmission. To address this issue, the integrated DE-MS-QP waveform is developed based on time-domain extension of the proposed MS-QP sequence, enabling simultaneous sensing and communication without mutual interference. Before we proceed, the following property is provided for DE-MS-QP waveform construction:

Let $\mathbf{s}_i=\left[s_i[0],s_i[1],\cdots,s_i[L_{s}-1]\right]$ ($i=0,1,\cdots,{M}'-1$) be $M'$ sequences of length $L_s$, and ${\mathbf{s}}'_i$ be the time-domain extension of $\mathbf{s}_i$ with length ${L}'_s=L_s{M}'$, expressed as
\begin{equation}\label{eq_lemma2_1}
{\mathbf{s}}'_i=\left[{s}'_i[0],{s}'_i[1],\cdots,{s}'_i[{L}'_s-1]\right]=\left[{\mathbf{s}}'_{i,0},{\mathbf{s}}'_{i,1},\cdots,{\mathbf{s}}'_{i,{M}'-1} \right]
\end{equation}
where ${\mathbf{s}}'_{i,g}=e^{\mathsf{j}\frac{2\pi gi}{{M}'}}{\mathbf{s}}_i$ for $g=0,1,\cdots,{M}'-1$.

Then the ${L}'_s$-DFT of ${\mathbf{s}}'_i$, denoted as ${S}'_i[k']$ for $k'=0,1,\cdots,{L}'_s-1$, can be calculated by
\begin{equation}\label{eq_lemma2_proof}
\begin{aligned}
{S}'_i[k']&=\sum_{n=0}^{{L}'_s-1}{s}'_i[n]\exp(-\textsf{j}\frac{2\pi k'n}{{L}'_s})\\&=\sum_{g=0}^{{M}'-1}\exp(-\textsf{j}\frac{2\pi \left (k'-i  \right )g}{{M}'})\sum_{n=0}^{L_s-1}s_i[n]\exp(-\textsf{j}\frac{2\pi k'n}{{L}'_s})\\&=\left\{\begin{matrix}
{M}'\sum_{n=0}^{L_s-1}s_i[n]\exp\left (-\textsf{j}\frac{2\pi k'n}{L_s'} \right ) , & \left \langle k' \right \rangle_{{M}'}=i; \\
0, & \text{else}.
\end{matrix}\right.
\end{aligned}
\end{equation}
Note that when $k'={M}'k+i$ with $k=0,1,\cdots,L_s-1$, ${S}'_i[k']$ can be further simplified as
\begin{equation}\label{eq_lemma2_4}
\begin{aligned}
S'_i[k']&=M'\sum_{n=0}^{L_s-1}s_i[n]\exp\left (-\textsf{j}\frac{2\pi (M'k+i)n}{L_s'} \right )\\&=M'\sum_{n=0}^{L_s-1}s_i[n]e^{-\textsf{j}\frac{2\pi in}{L'_s}}\exp\left (-\textsf{j}\frac{2\pi kn}{L_s} \right ),
\end{aligned}
\end{equation}
which can be seen as the $k$-th scaled $L_s$-DFT coefficient of a phase-shifted version of $\mathbf{s}_i$, denoted by $\bar{\mathbf{s}}_i=\left [ \bar{s}_i[0],\bar{s}_i[1],\cdots,\bar{s}_i[L_s-1] \right ]$ with $\bar{s}_i[n]=s_i[n]e^{-\textsf{j}\frac{2\pi in}{{L}'_s}}$.

It is seen from above that time-domain extension of the transmitted sequences enables multiple concurrent data streams on non-intersect frequency points without mutual interference. Following this philosophy, the proposed DE-MS-QP waveform is constructed as Fig. \ref{fig6}. For the $m$-th subband ($m=0,1,\cdots,M-1$), the incoming ZC sequence $\mathbf{b}_m$ is repeated for ${M}'$ times, yielding ${\mathbf{b}}'_m$ of length ${L}'_m=L_m{M}'$ formulated as
\begin{equation}
\begin{aligned}
{\mathbf{b}}'_m=\underset{{M}'\text{ times}}{\underbrace{\left[\mathbf{b}_m,\mathbf{b}_m,\cdots,\mathbf{b}_m\right]}}=\left [ {b}'_m[0],{b}'_m[1],\cdots,{b}'_m[{L}'_m-1] \right ],
\end{aligned}
\end{equation}
whose ${L}'_m$-DFT coefficients are denoted as ${B}'_m[k']$ ($k'=0,1,\cdots,L_m{M}'-1$). According to (\ref{eq_lemma2_1})-(\ref{eq_lemma2_proof}), the extended ZC sequence only occupies the ${M}'k$-th frequency points for $k=0,1,\cdots,L_m-1$, in other words, ${B}'_m[k']=0$ for $\langle k' \rangle_{{M}'}\neq 0$.

\begin{figure*}[t!]
\begin{center}
\includegraphics[width=0.9\linewidth, keepaspectratio]{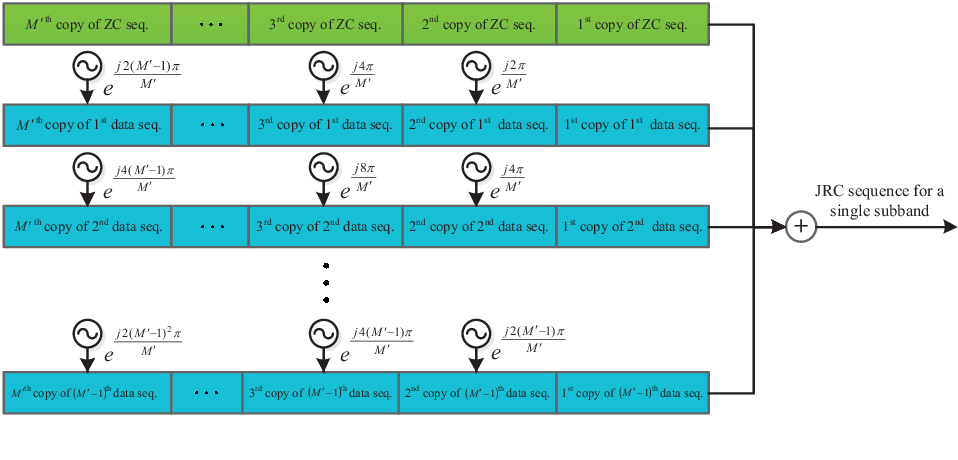}
\end{center}
\caption{Diagram of sequence generation for each subband of the proposed DE-MS-QP waveform.}
\label{fig6}
\end{figure*}
The resultant zero frequency points can be utilized for data transmission. As presented in Fig. \ref{fig6}, totally $({M}'-1)$ $L_m$-length data sequences modulated with the constellation alphabet $\mathcal{M}$, expressed by $\mathbf{s}_{i,m}=\left[s_{i,m}[0],s_{i,m}[1],\cdots,s_{i,m}[L_m-1] \right]$ for $i=1,2,\cdots,M'-1$, are periodically transmitted for ${M}'$ times. Meanwhile, phase rotation is imposed on each copy of the data sequence, yielding the extended data sequence written as
\begin{equation}
\begin{aligned}
{\mathbf{s}}'_{i,m}&=\left [ {\mathbf{s}}'_{i,m,0}, {\mathbf{s}}'_{i,m,1},\cdots,{\mathbf{s}}'_{i,m,{M}'-1} \right ]\\&=\left[{s}'_{i,m}[0],{s}'_{i,m}[1],\cdots,{s}'_{i,m}[{L}'_m-1]\right],
\end{aligned}
\end{equation}
where we have
\begin{equation}
{\mathbf{s}}'_{i,m,g}=e^{\textsf{j}\frac{2\pi gi}{{M}'}}\mathbf{s}_{i,m}.
\end{equation}
According to (\ref{eq_lemma2_1})-(\ref{eq_lemma2_proof}), ${\mathbf{s}}'_{i,m}$ merely takes up the $({M}'k+i)$-th frequency points ($k=0,1,\cdots,L_m-1$) by performing an $L'_m$-DFT for $i=1,\cdots,M'-1$, respectively. Therefore, the extended sensing and data sequences can be superposed together without mutual interference, written as
\begin{equation}
{x}'_m[n]={b}'_{m}[n]+\sum_{i=1}^{{M}'-1}{s}'_{i,m}[n],\:\: n=0,1,\cdots,{L}'_m-1,
\end{equation}
which is then moved to the $m$-th subband utilizing the transmitter structure in Fig. \ref{fig3} for $m=0,1,\cdots,M-1$, respectively. Finally, the proposed DE-MS-QP waveform is obtained by adding up resultant signals on each subband, whose equivalent baseband expression is formulated as
\begin{equation}\label{eq_de_msqp}
\begin{aligned}
{x}'[n]&\overset{\triangle }{=}\frac{1}{\sqrt{{N}'}}\sum_{m=0}^{M-1}e^{\textsf{j}\left(\frac{2\pi f'_mn}{{N}'}+\hat{\phi}_m\right)}\frac{1}{\sqrt{{L}'_m}}\sum_{l=0}^{{L}'_m-1}{x}'_m[l]\times\\&\frac{\sin\left ( {L}'_m\pi(\frac{n}{{N}'}-\frac{l}{{L}'_m}) \right )}{\sin\left ( \pi(\frac{n}{{N}'}-\frac{l}{{L}'_m}) \right )} e^{\textsf{j}\pi({L}'_m-1)(\frac{n}{{N}'}-\frac{l}{{L}'_m})},
\end{aligned}
\end{equation}
where ${N}'=(\sum_{m=0}^{M-1}{L}'_m)+M{L}'_G$ denotes the length of ${x}'[n]$, and ${L}'_G$ represents the length of GI in the frequency domain. Besides, $f'_m$ is defined as
\begin{equation}
f'_m=\left\{\begin{matrix}
(\sum_{i=0}^{m-1}L'_{i})+mL'_G, & 1\leq m\leq M-1;\\
0, & m=0.
\end{matrix}\right.
\end{equation}
The proposed DE-MS-QP waveform shares the same multi-subband structure as the proposed MS-QP sequence, which could achieve ultra-high-resolution ranging with cost-efficient front-end devices. Besides, its radar sensing component can be seen as periodical transmissions of the MS-QP sequence for ${M}'$ times, which is uncorrelated with the data component since they occupy non-intersect frequency points. Therefore, the DE-MS-QP waveform not only ensures desirable sensing performances inherited from the MS-QP sequence, but also supports simultaneous data transmission without interference from the radar counterpart theoretically. Its communication spectral efficiency can be formulated as
\begin{equation}\label{eq_se_cal}
\text{SE}=\frac{{N}'-ML'_G}{{N}'+L_{\text{cp}}}\times\frac{{M}'-1}{{M}'}\log_2\left | \mathcal{M} \right |\:\:(\text{bit/s/Hz}),
\end{equation}
which considers the insertion of cyclic prefix (CP) of length $L_{\text{cp}}$ against the timing error or inter-symbol interference (ISI).

{\bf Remark 2}: There is a trade-off in frequency resource allocation between radar sensing and communication applications. More specifically, with the increase of ${M}'$, more independent data symbol streams are transmitted in parallel, leading to enhanced throughput at the cost of reduced power supply for radar sensing, which {may cause} degradation of the sensing performance. {Explicitly, we assume equal power allocation for $M'$ different radar sensing or communication streams for simplicity. Then the receiving SNR corresponding to the radar sensing component of the proposed DE-MS-QP waveform can be formulated as
\begin{equation}
	\text{SNR}_{\text{radar}}=10\log_{10}P_{r}/(M'P_w),	
\end{equation}
where $P_r$ and $P_w$ denote the average power of the echo signals and the noise components at the sensing receiver, respectively. It is seen that $\text{SNR}_{\text{radar}}$ is degraded with the increase of $M'$.}

{To elaborate a little further, with the increase of $\text{SNR}_{\text{radar}}$, the range and velocity estimation errors will finally converge to the sensing resolution at a threshold SNR value. When $\text{SNR}_{\text{radar}}$ exceeds this threshold, it is expected that the value of $M'$ can be enlarged for higher communication spectral efficiency, without causing any degradation of the sensing performance.} The detailed investigation will be provided in Section \ref{s5}.
\section{Receiver Processing Techniques}\label{s4}
\subsection{Receiver Design for Radar Sensing with the Proposed MS-QP Sequence}\label{s4.1}
For radar sensing applications using the proposed MS-QP sequence, which is generated by the transmitter architecture in Fig. \ref{fig3}, inverse operations can be performed at the receiver as shown in Fig. \ref{fig7}. Firstly, the received echo signal is down-converted and filtered to extract the signal components on each subband, which are then sampled by cost-efficient A/Ds with low sampling rate of $1/T_m$ for $m=0,1,\cdots,M-1$, respectively, leading to reduced hardware cost without the need of the full-band A/D with sampling rate over tens of GHz. Afterwards, following the principle of MS-QP sequence construction introduced in Section \ref{s3.1}, the outputs of the $M$ subbands are utilized {by the digital signal processing (DSP) module} to re-construct the digital form of the echoes for radar sensing. {To elaborate more, in practical, the coherent processing interval (CPI) for radar sensing needs to be sufficiently long to enhance the SNR at the radar receiver against the severe path loss, which is assumed to contain $Q$ consecutive sequence transmissions. Considering the $q$-th $N$-length transmission subblock, the outputs by the $m$-th A/D, denoted as $\mathbf{b}_{r,q,m}=[b_{r,q,m}[0],b_{r,q,m}[1],\cdots,b_{r,q,m}[L_m-1]]$, can be expressed as
	\begin{equation}\label{eq_revision_1}
		b_{r,q,m}[n]\approx\sum_{i=1}^{I}\left(b_m[n-\left \lfloor\frac{T_m}{T_s}\tau_i \right \rfloor]e^{\textsf{j}\frac{2\pi v_inT_m}{T_s}}\right)+w_{q,m}[n],
	\end{equation}
where $w_{q,m}[n]$ is the noise term, and hardware imperfection components are omitted for brevity. $\mathbf{b}_{r,m}$ for $m=0,1,\cdots,M-1$ are fed into the DSP module, and each performs an $L_m$-point DFT to generate $\mathbf{B}_{r,q,m}$, respectively. Then $\mathbf{B}_{r,q,m}$ for $m=0,1,\cdots,M-1$ are concatenated together with the $L_G$-length guard interval inserted between any two neighbouring subbands, written as
\begin{equation}
	\mathbf{Y}_q=\left [ \mathbf{B}_{r,q,0},\mathbf{0},\mathbf{B}_{r,q,1},\mathbf{0},\cdots,\mathbf{B}_{r,q,M-1},\mathbf{0} \right],	
\end{equation}
where $\mathbf{Y}_q$ is the frequency-domain echo signal of length $N$, and $\mathbf{0}$ denotes the $L_G$-length all-zero vector. Finally, the $q$-th received subblock for $q=0,1,\cdots,Q-1$, denoted as $\mathbf{y}_q=\left[y_q[0],y_q[1],\cdots,y_q[N-1]\right]$, could be obtained by taking an $N$-point IDFT of $\mathbf{Y}_q$, expressed as
\begin{equation}
	y_q[n]\approx\sum_{i=1}^{I}h_i\mu _rx[n-\tau_i]e^{\textsf{j}(2\pi (n+qN){v}_i)}+\tilde{w}_q[n],
\end{equation}
where $\tilde{w}_q[n]$ denotes the noise component. }The received subblocks are then used for target detection via range-Doppler-matrix-based (RDM-based) algorithms \cite{Zeng_tvt_20}. As illustrated in Fig. \ref{fig8}, the cyclic cross-correlation between each received subblock and the transmitted MS-QP sequence is firstly calculated as
\begin{figure}[t!]
	\begin{center}
		\includegraphics[width=0.9\linewidth, keepaspectratio]{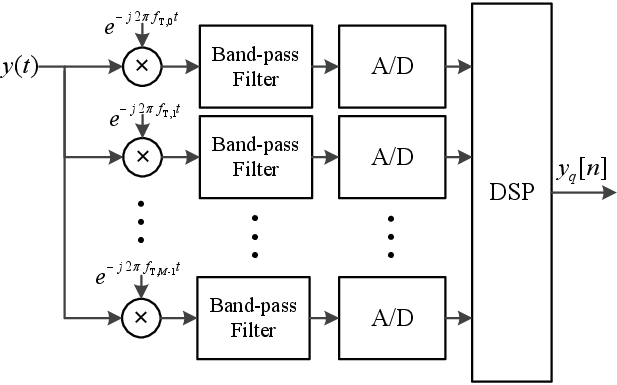}
	\end{center}
	\caption{Receiver diagram for MS-QP sequence collection. }	
	\label{fig7}
\end{figure}
\begin{figure*}[t!]
\begin{center}
\includegraphics[width=0.9\linewidth, keepaspectratio]{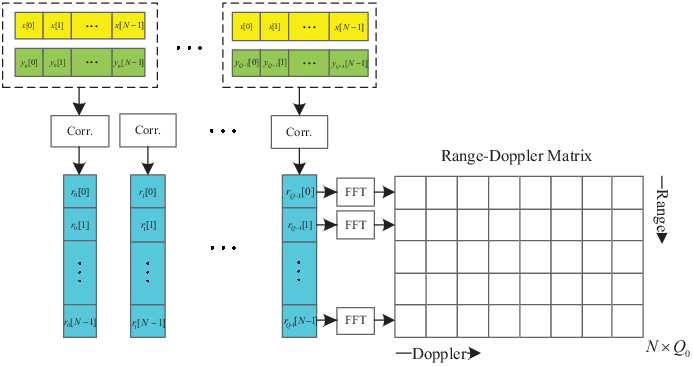}
\end{center}
\caption{Receiver processing techniques for radar sensing, where ``Corr.'' is referred to as the cyclic correlation operation.}
\label{fig8}
\end{figure*}
\begin{equation}
r_q[n]=\sum_{i=0}^{N-1}y_q[i]x^*[i-n],\:\:n=0,1,\cdots,N-1.
\end{equation}
Then an FFT with size of $Q_0=wQ$ ($w$ is an arbitrary positive integer) is performed on the correlation values at time instant $n$, i.e., $r_q[n]$ for $q=0,1,\cdots,Q-1$, formulated as
\begin{equation}
R(n,k)=\sum_{q=0}^{Q-1}r_{q}[n]e^{-\textsf{j}\frac{2\pi qk}{Q_0}},
\end{equation}
which is defined as the $(n,k)$-th RDM element for $n=0,1,\cdots,N-1$ and $k=0,1,\cdots,Q_0-1$. Afterwards, $R(n,k)$ is utilized for target detection via constant false alarm rate (CFAR) approaches \cite{Kronauge_taes_13}, written as
\begin{equation}
\text{H}_1:\frac{\left \| R(n,k) \right \|^2}{\hat{\lambda}(n,k)}\geq\Gamma \:\: \text{H}_0: \frac{\left \| R(n,k) \right \|^2}{\hat{\lambda}(n,k)}<\Gamma,
\end{equation}
where $\text{H}_1$ and $\text{H}_0$ represent the hypothesis for the presence or absence of a target at the $(n,k)$-th cell, respectively. $\hat{\lambda}(n,k)$ and $\Gamma$ denote the estimated average noise floor at the $(n,k)$-th cell and the decision threshold. Readers are referred to \cite{Zeng_tvt_20} and references therein for more details.

In the mmWave and low-THz bands, strong Doppler shift is likely to induce high sidelobes on the range profile, which could be wrongly detected as targets by the classical CFAR approach, if satisfying the hypothesis $\text{H}_1$. To address this issue, {we notice that different targets are unlikely to be located too closely in practical, which indicates that there hardly exists other target within the neighbourhood of the already detected target corresponding to the main peak. This inspires us to move the dominant sidelobes caused by the Doppler shift into the neighbourhood of the main peak, and then we could exclude them from the real targets, leading to reduction of the possible false alarms. More specifically,} the root indices of $\mathbf{b}_m$ ($m=0,1,\cdots,M-1$) are set as $p_m\in\{1,\frac{L_m-1}{2},\frac{L_m+1}{2},L_m-1\}$ to concentrate the dominant sidelobes closely around the main peaks on the range profile. Then a target detection criterion, named as Target Exclusion nearby the Main Peak (TEMP), is proposed by assuming no other reflector nearby the detected target corresponding to the main peak at $R(\hat{n},\hat{k})$, i.e., no target within the interval $\left [ \hat{n}- \bar{n},\hat{n} \right )\cup \left (\hat{n},\hat{n}+ \bar{n} \right ]$. Here $\bar{n}$ is a positive integer parameter determined empirically. By employing the proposed TEMP strategy, the procedure of target detection is modified as Algorithm \ref{alg1}.

After target detection, the range and velocity estimates of the targets corresponding to $\mathcal{I}=\{(\hat{n}_0,k_{\hat{n}_0}),(\hat{n}_1,k_{\hat{n}_1}),\cdots,(\hat{n}_{I-1},k_{\hat{n}_{I-1}})\}$ obtained in Algorithm \ref{alg1}, denoted as $\hat{d}_i$ and $\hat{u}_i$ for $i=0,1,\cdots,I-1$, can be calculated by \cite{Zeng_tvt_20}
\begin{equation}
\hat{d}_i=\frac{c_0\hat{n}_iT_s}{2},\:\:
\end{equation}
\begin{equation}
\hat{u}_i=\left\{\begin{matrix}
{c_0\hat{k}_{\hat{n}_i}}/(2Q_0Nf_cT_s), &\hat{k}_{\hat{n}_i}<\frac{Q_0}{2};  \\
{c_0(\hat{k}_{\hat{n}_i}-Q_0)}/(2Q_0Nf_cT_s), & \hat{k}_{\hat{n}_i}\geq\frac{Q_0}{2}.
\end{matrix}\right.
\end{equation}

\begin{algorithm}[t!]
\caption{Target Detection Algorithm based on the Proposed TEMP Strategy}
\label{alg1}
\begin{algorithmic}[1]
\Require
 The RDM elements $R(n,k)$, Estimates of the average noise floor at the $(k,m)$-th cell $\hat{\lambda}(k,m)$, the decision threshold $\Gamma$, the positive integer parameter $Q_0$, $\bar{n}$ and $N$;
\Ensure
 The set $\mathcal{I}$ containing RDM coordinates of the detected targets determining their corresponding round-trip delays and Doppler shifts;
 \State $\mathcal{I}=\varnothing$;
 \For {($n=0;n\leq N-1;n++$)}
 \State $\hat{k}_n=\arg\underset{0\leq k_n\leq {Q}_0-1}{\max} \left \| R(n,k_n) \right \|$;
 \EndFor
 \State $m=0$;
 \State $\mathcal{N}^{(0)}=\left \{0,1,\cdots,N-1\right \}$;
 \Repeat
    \State $\hat{n}^{(m)}=\arg\underset{n\in\mathcal{N}^{(m)}}{\max} \left \| R(n,\hat{k}_n) \right \|$;
    \If{$\frac{\left \| R(\hat{n}^{(m)},\hat{k}_{\hat{n}^{(m)}}) \right\|^2}{\hat{\lambda}(\hat{n}^{(m)},\hat{k}_{\hat{n}^{(m)}})}\geq\Gamma$}
    \State $\mathcal{I}=\mathcal{I}\vee \{(\hat{n}^{(m)},\hat{k}_{\hat{n}^{(m)}})\}$;
    \State $\mathcal{N}^-=\left \{ n\mid n\in[\hat{n}^{(m)}-\bar{n},\hat{n}^{(m)}+\bar{n}]\right \}$;
    \State $\mathcal{N}^{(m+1)}=\mathcal{N}^{(m)}\setminus \mathcal{N}^-$;
    \Else
    \State $\mathcal{N}^{(m+1)}=\mathcal{N}^{(m)}\setminus \{\hat{n}^{(m)}\}$;
    \EndIf
    \State $m++$
 \Until{($\mathcal{N}^{(m)}=\varnothing$)}
\State \Return $\mathcal{I}$;

\end{algorithmic}
\end{algorithm}

\subsection{Receiver Design for JRC Applications with the Proposed DE-MS-QP Waveform}
Similar procedures can be employed to process the received DE-MS-QP signals for radar sensing, including low-rate A/D operations on each subbands and correlation-based RDM calculation, etc. In RDM calculation, one straightforward way is to directly perform the cyclic correlations of length ${N}'={M}'N$ between the received signals and the extended MS-QP sequence. However, it suffers from large computational complexity for cyclic correlations, calculated as $\mathcal{O}(\left ({M}'N  \right )^2)$ in terms of complex multiplications, and cannot support flexible adjustment of ${M}'$ since the correlation length changes with $M'$. To this end, each received DE-MS-QP frame is divided into $M'$ length-$N$ subblocks, where cyclic correlation of length $N$ between each subblock and the MS-QP sequence is directly applied. These correlation results are then utilized for radar sensing as described in Section \ref{s4.1}, which is omitted for brevity. This proposed enhancement not only reduces the correlation complexity of the radar receiver to $\mathcal{O}({M}'N^2)$, but enables flexible parameter adjustment as well. Moreover, the cross-correlation between data components of the echoes and the MS-QP sequence is usually small when $N$ is sufficiently large, as validated in \cite{mao_supportingfile_21}, thus posing marginal impact on radar sensing.

On the other hand, the data symbols can be demodulated at the communication receiver without interference from sensing sequences, as presented in Fig. \ref{fig9}. Explicitly, we assume that each CPI contains $Q'$ times consecutive transmission of the DE-MS-QP waveform. Then the frequency-domain components on the $m$-th subband of the $q$-th received subblock can be obtained as $\mathbf{Y}'_{m,q}=\left[{Y}'_{m,q}[0],{Y}'_{m,q}[1],\cdots,{Y}'_{m,q}[L_m'-1]\right]$ for $q=0,1,\cdots,Q'-1$ and $m=0,1,\cdots,M-1$. The radar sensing component is firstly extracted as $\mathbf{Y}'_{0,m,q}=\left[{Y}'_{m,q}[0],{Y}'_{m,q}[{M}'],\cdots,{Y}'_{m,q}[(L_m-1){M}']\right]$, which could be employed for channel estimation to save additional pilot overhead. With the knowledge of channel state information, the data components of the $({M}'-1)$ data streams, denoted by $\mathbf{Y}'_{i,m,q}=\left[{Y}'_{m,q}[i],{Y}'_{m,q}[{M}'+i],\cdots,{Y}'_{m,q}[(L_m-1){M}'+i]\right]$ for $i=1,2,\cdots,M'-1$, are equalized, and then transformed to time domain using $L_m$-IFFT operations, yielding $\mathbf{y}'_{i,m,q}=\left[{y}'_{i,m,q}[0],{y}'_{i,m,q}[1],\cdots,{y}'_{i,m,q}[L_m-1]\right]$, respectively. According to (\ref{eq_lemma2_4}) and (\ref{eq_de_msqp}), phase adjustment is performed on each element of $\mathbf{y}'_{i,m,q}$ for $i=1,2,\cdots,M'-1$, written as
\begin{equation}
{y}''_{i,m,q}[n]={y}'_{i,m,q}[n]\times e^{\textsf{j} (\frac{2\pi in}{L'_m} -\hat{\phi}_m )},\:\: n=0,1,\cdots,L_m-1,
\end{equation}
which is finally demodulated with a maximum-likelihood (ML) detector.

\begin{figure}[t!]
	\begin{center}
		\includegraphics[width=0.9\linewidth, keepaspectratio]{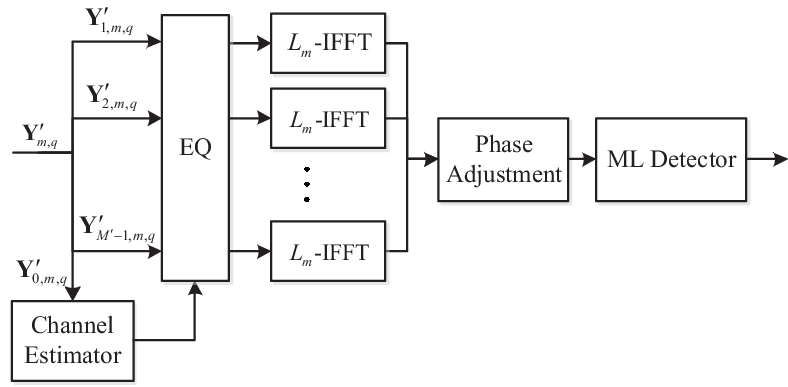}
	\end{center}
	\caption{Demodulator diagram for each subband of the proposed DE-MS-QP waveform at the communication receiver. The frequency-domain equalizer is abbreviated as ``EQ''.}
	\label{fig9}
\end{figure}
\section{Numerical Results}\label{s5}
In this section, {a sub-mmWave JRC system with the carrier frequency of 240 GHz is considered.} We firstly evaluate the feasibility of proposed root index design for the MS-QP sequence together with TEMP strategy in radar sensing. Afterwards, the accuracy of target ranging and velocity estimation is compared between the proposed waveforms and the classical counterparts, including ZC sequences \cite{AlSharif_conf_17} and LFM signals \cite{Patole_spm_17}. Moreover, the performance trade-off between radar sensing and communication of the proposed DE-MS-QP waveform is also explored numerically. In simulations, the symbol rate of different candidate sequences is assumed to be equal to the transmission bandwidth according to the Nyquist theory. Furthermore, in order to simulate the fractional time delay and shaping filtering, $4$-times upsampling is performed on the transmitted signals, where the round-trip delay and low-pass filtering are imposed at the higher sampling rate. On the other hand, the echo signals at the receiver will be down-sampled to the original symbol rate for further radar detection \cite{Zeng_tvt_20}. Besides, for hardware imperfections, the random variation term for phase noise $\Delta\theta_n$ follows Gaussian distribution written as $\mathcal{N}(0,(0.3^{\circ})^2)$, and the RX amplitude and phase imbalances are set as $\epsilon_r=0.2$ and $\phi_r=10^{\circ}$, respectively \cite{Ramadan_access_18}. {Finally, we define ``SNR'' in simulation results as the receiving SNR corresponding to the echo signal, calculated as the ratio between the power of echoes and noise at the radar receiver.}

Figure \ref{fig10} illustrates the false alarm rate of radar sensing with respect to $\Gamma$ using the proposed MS-QP sequence and the TEMP detection strategy with {$\bar{n}=20$}, with/without the proposed root index design against Doppler shift. In simulations, the SNR at the radar receiver is set as $-40$ dB. Besides, we consider totally $3$ targets with velocity of $20$ m/s, randomly distributed in $[0, 10]$ m from the JRC platform, where the distances between neighbouring targets are assumed to be no less than $0.3$ m. Moreover, the number of subbands for the proposed MS-QP sequence is set as $M=10$. The lengths of GIs between adjacent subbands and the ZC subsequences are set as $L_G=100$ and $L_m=10007$ for $m=0,1,\cdots,9$, respectively. Here we assume that the MS-QP sequence is repetitively transmitted for $Q=100$ times in each CPI. Then it is seen in Fig. \ref{fig10} that, when $p_m=3$ for different ZC subsequences, there exists a severe floor of the false alarm rate as $\Gamma$ increases, which is induced by the dominant range sidelobes from Doppler shift. On the other hand, with proposed root index design $p_m\in\{5003,5004,1,10006\}$, the probability of false alarms declines rapidly with the increase of $\Gamma$, all attaining significant performance gain of target detection over their counterpart of $p_m=3$. This shows superior robustness of the MS-QP sequence with proposed root index design against Doppler shift effects. Furthermore, $p_m\in\{1,10006\}$ achieves lower false alarm rate than $p_m\in\{5003,5004\}$, since the former concentrates the dominant sidelobes closer to the main peak than the latter.

%

\begin{figure}[t!]
	\begin{center}
		\includegraphics[width=0.9\linewidth, keepaspectratio]{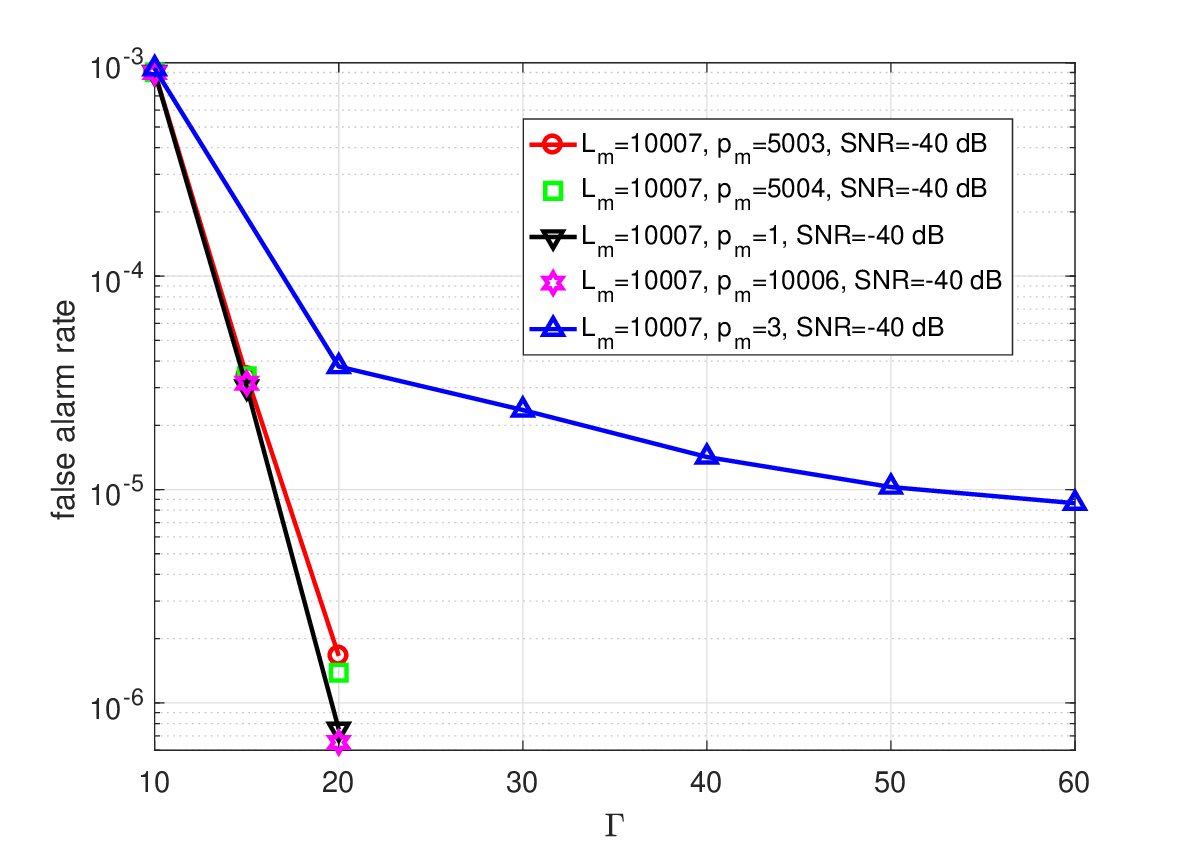}
	\end{center}
	\caption{False alarm rate of radar sensing using MS-QP sequences with/without the proposed root index design, where the proposed TEMP strategy is applied for target detection in both cases.}
	\label{fig10}
\end{figure}

\begin{figure*}[t!]
	\centering
	\subfigure[Ranging performance.]{
		\label{fig11} 
		\includegraphics[width=0.4\linewidth]{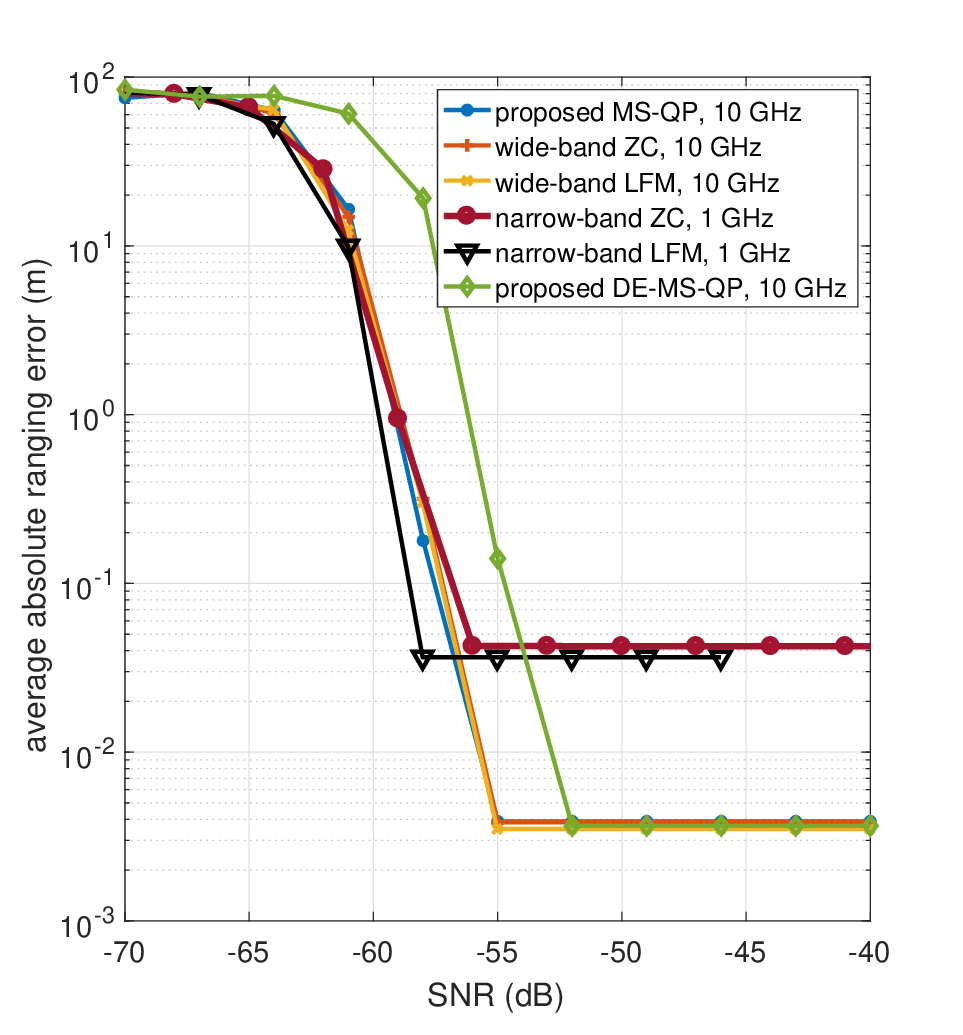}}
	\subfigure[Velocity estimation performance.]{
		\label{fig12} 
		\includegraphics[width=0.4\linewidth]{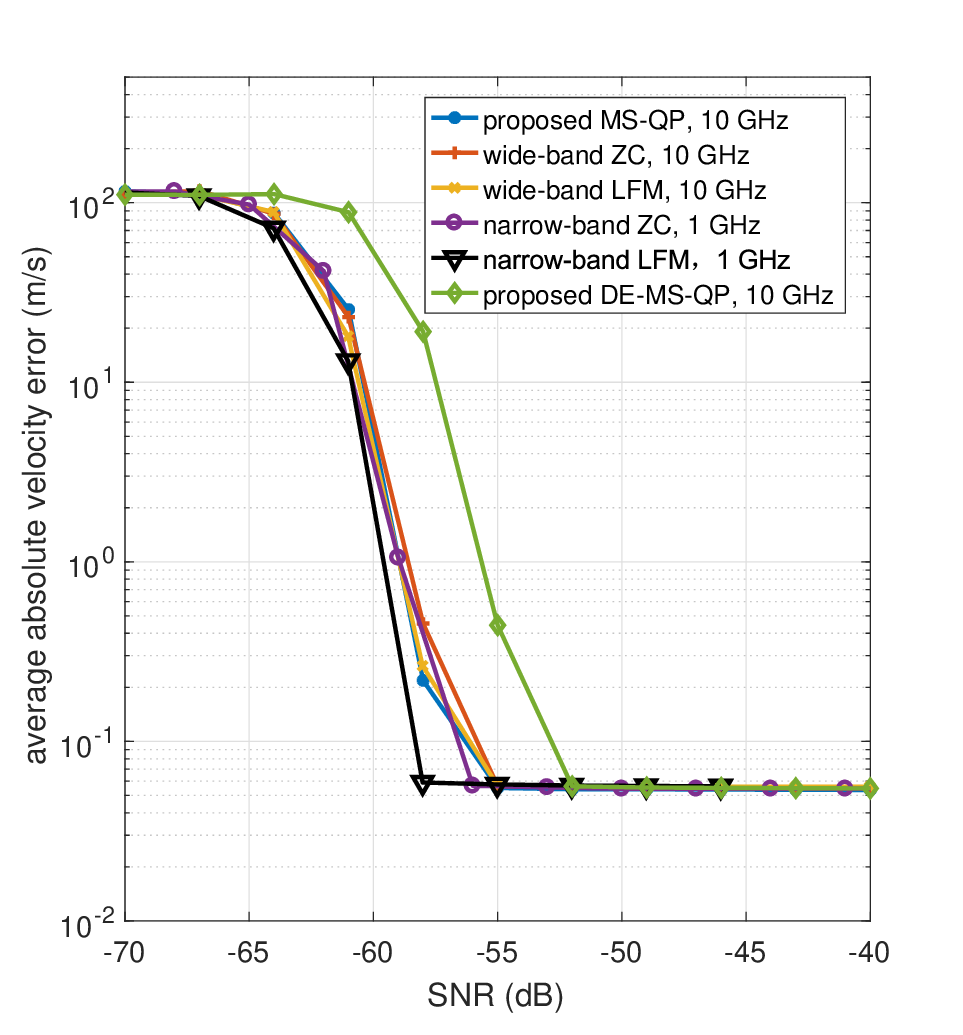}}
	\caption{Performance comparison of target ranging and velocity estimation between the proposed waveforms and their classical counterparts, where SNR is referred to as the signal-to-noise ratio at the radar receiver.}
\end{figure*}

Figures \ref{fig11} and \ref{fig12} present the accuracy comparison of the target ranging and velocity estimation between the proposed MS-QP sequence and its classical counterparts. In simulations, the target range and relative velocity are randomly distributed in $[0, 3]$ m and $[-20, 20]$ m/s, respectively. For different waveforms, each CPI contains $Q=1024$ periodical sequence transmissions, and the duration per sequence transmission is fixed as $1.1077$ $\mu$s. An MS-QP sequence of $N=11077$ is considered, which is constituted by $M=10$ identical ZC sequences with the length and root index equal to $1007$ and $503$, on different subbands divided by GIs of $L_G=100$, taking up $10$-GHz bandwidth in total. On the other hand, ZC sequences and LFM signals are also considered for radar sensing: 1) a narrow-band ZC sequence of $1$ GHz bandwidth, with the length and root index equalling $1007$ and $503$; 2) a wide-band ZC sequence of $10$ GHz bandwidth, with the length and root index equalling $11077$ and $5538$; {3) an LFM waveform of $1$-GHz bandwidth, with pulse length of $1.1077$ $\mu$s;} 4) an LFM waveform of $10$-GHz bandwidth, with pulse length of $1.1077$ $\mu$s. It can be observed in Fig. \ref{fig11} that, the MS-QP sequence is capable of significantly reducing the average ranging error by about $20$ dB compared with the narrow-band ZC sequence and {LFM signals}, both using $1$-GHz A/Ds at the SNRs over $-55$ dB. Besides, as is illustrated in Figs. \ref{fig11} and \ref{fig12}, in comparison with the wide-band ZC sequences and LFM signals requiring {a full-band} 10-GHz A/D, the proposed MS-QP sequence is capable of achieving the same estimation error of target range and velocity, obtained as $0.004$ m and $0.06$ m/s over $-55$-dB SNRs, with {comparable radar processing complexity and the only need of several cost-efficient 1-GHz A/Ds.}. Although the constant-amplitude property is lost compared with its classical counterparts, the proposed MS-QP sequence manages to attain an acceptable level of PAPR as $6.6$ dB with the aid of phase rotation on different subbands, achieving $7$ dB PAPR reduction than that without phase rotation, calculated as $13.7$ dB. Additionally, the sensing performance of the proposed DE-MS-QP waveform with $M'=2$ is also simulated, which is constructed based on the MS-QP sequence considered above, and utilizes QPSK for modulation. Besides, $Q'=Q/M'=512$ consecutive DE-MS-QP frames are contained in each CPI, with CP added at the beginning of each frame, whose length is set as $1$\% of the DE-MS-QP frame. Although the received SNR of the sensing sequences is degraded by $3$ dB thanks to the embedded data consuming half of the transmit power, the proposed DE-MS-QP waveform is still capable of attaining low estimation errors of target range and velocity, i.e., $0.004$ m and $0.06$ m/s at SNRs over $-52$ dB. Meanwhile, it supports high-rate data transmission of about $10$ Gb/s.

%
\begin{figure*}[t!]
	\centering
	\subfigure[Ranging performance.]{
		\label{fig13} 
		\includegraphics[width=0.4\linewidth]{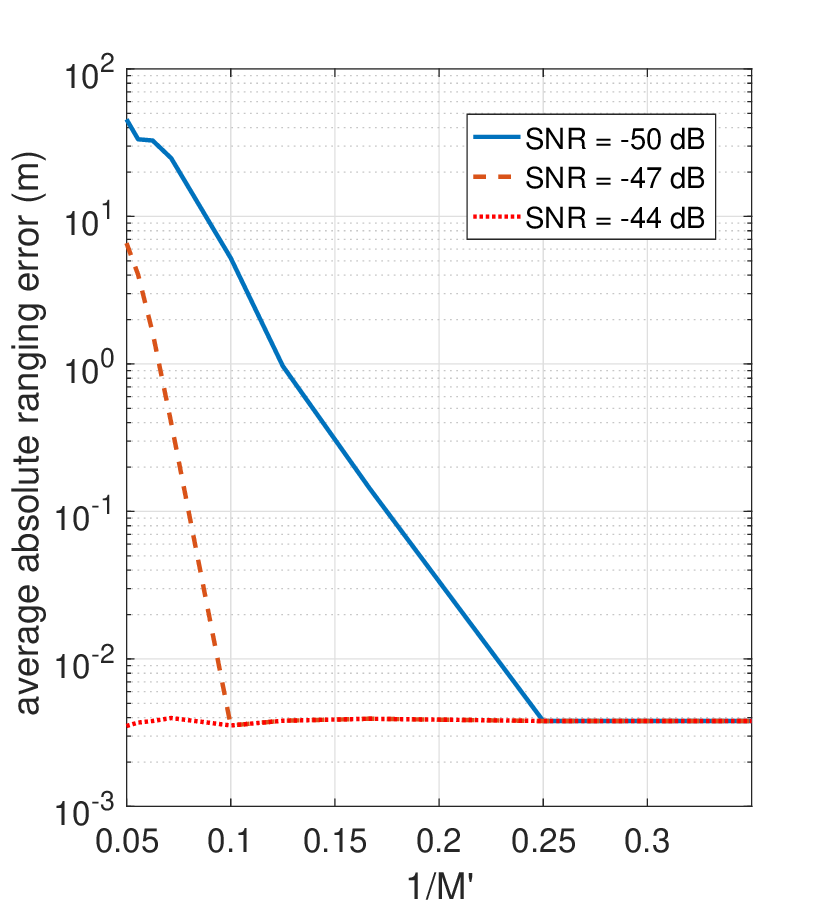}}
	\subfigure[Velocity estimation performance.]{
		\label{fig14} 
		\includegraphics[width=0.4\linewidth]{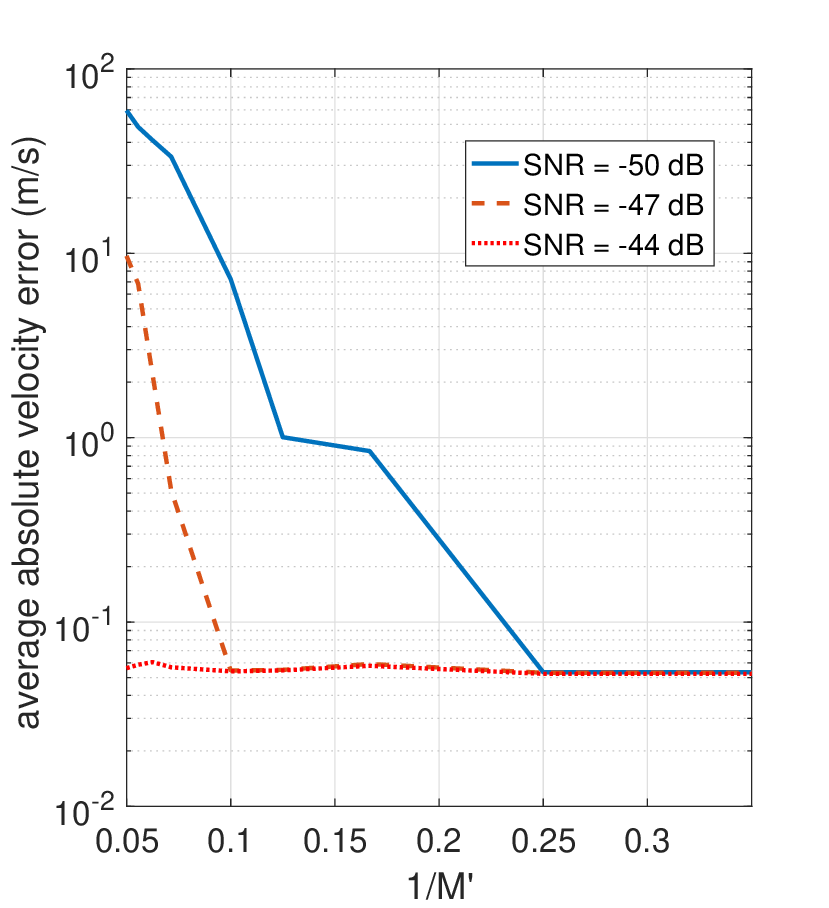}}
	\caption{Sensing performances of the DE-MS-QP waveform with respect to $1/M'$ under different SNRs at the radar receiver.}
\end{figure*}

Figures \ref{fig13} and \ref{fig14} illustrate the sensing performance of the proposed DE-MS-QP waveform with respect to $1/M'$, which denotes the proportion of frequency resources distributed for radar sensing. In the simulation, the parameter settings are the same as those in Figs. \ref{fig11} and \ref{fig12} except for the value of $M'$. {Firstly, according to (\ref{eq_se_cal}), the communication spectral efficiency gradually improves with the increase of $M'$ (or decrease of $1/M'$).} On the other hand, it is observed that, at the SNRs of $-50$ and $-47$ dB, the sensing performance is degraded with the decrease of $1/M'$ when $1/M'$ is smaller than $0.25$ and $0.1$, respectively, despite the enhanced communication spectral efficiency. This is due to the reduction of power supply for the sensing application. On the other hand, when $1/M'$ is larger than $0.25$ and $0.1$ for SNR values of $-50$ and $-47$ dB, the estimation errors of target range and velocity will not deteriorate with the decrease of $1/M'$, remaining to be below $0.01$ m and $0.1$ m/s, respectively. Hence, it is concluded that there exists a unique threshold of $M'$ (or $1/M'$) for different received SNRs, i.e., $4$ and $10$ for $-50$ and $-47$ dB SNRs. This threshold becomes larger with the increase of the received SNR, and even reaches over $20$ at the SNR of $-44$ dB. The performance of radar sensing will not be affected by the enlargement of $M'$ until exceeding the threshold. Therefore, $M'$ could be set as this threshold value (determined numerically) in order to realize the best trade-off of frequency resource allocation, guaranteeing superior sensing performance and data throughput simultaneously. For instance, when the received SNR equals $-47$ dB, the proposed DE-MS-QP sequence is capable of achieving accurate target ranging and velocity estimation with only $1/10$ of the available frequency resources, whilst the rest can be employed for high-rate data transmission, which approaches 16 Gb/s with QPSK modulation.

Finally, Fig. \ref{fig15} presents the bit-error rate (BER) performance of the proposed DE-MS-QP waveform versus receiving SNR at the user side, which considers different values of $M'$. In the simulation, we assume perfect channel estimation and phase tracking for the communication subsystem. It is seen that, the data throughput is improved with the increase of $M'$, at the cost of BER performance degradation. For instance, the proposed DE-MS-QP waveform could enhance 0.72 bit/s/Hz spectral efficiency by tuning $M'$ from $2$ to $10$, whilst leading to $1$ dB performance loss at the BER of $10^{-3}$ at the same time. This shows an interesting trade-off between the BER performance and the data throughput of the DE-MS-QP waveform. {Besides, compared with classical OFDM, there is about $1$ dB performance loss at the BER of $10^{-3}$ of the proposed DE-MS-QP waveform compared with classical OFDM, at the spectral efficiency of $1.6$ bit/s/Hz. However, the OFDM waveform inherently suffers from high PAPR issue, causing severe nonlinear distortions especially for mmWave/THz-band power amplifiers. Moreover, unlike the proposed DE-MS-QP waveform, a full-band A/D of $10$-GHz sampling rate is required at the OFDM receiver, leading to higher hardware cost. By considering the merits over OFDM, the slight performance loss of the proposed DE-MS-QP waveform is acceptable, which is more suitable for JRC applications in the mmWave/THz scale.}

\section{Conclusions and Future Prospects}\label{s6}
To overcome the technical challenges of radar sensing and communication convergence at mmWave and low-THz frequencies, the waveform design issue is investigated in this paper. For TDD-based JRC systems, we propose the MS-QP sequence with broad bandwidth for radar sensing, which incorporates multiple ZC sequences on different subbands in the mmWave/THz scale. The proposed MS-QP sequence could achieve ultra-high-resolution sensing only with cost-efficient A/Ds with low sampling rate, whose transceiver structure is also compatible with state-of-the-art subband filtered communication systems. Furthermore, to mitigate strong Doppler shift effects on target detection, the root indices of ZC subsequences on different subbands of the MS-QP sequence is designed to concentrate the dominant sidelobes closely around the main peak on the range profile. Then false alarms induced by Doppler shifts can be eliminated by further invoking the TEMP strategy. Moreover, to support simultaneous radar sensing and data transmission, the DE-MS-QP waveform is developed based on time-domain extension of the MS-QP sequence, where the sensing and data components are modulated on non-intersect frequency points, thus avoiding mutual interference.

Numerical results demonstrate the superiority of the proposed root index design for MS-QP sequences in terms of the false alarm rate, and that the proposed MS-QP and DE-MS-QP waveforms are capable of attaining ultra-high-resolution ranging and velocity estimation under extremely noisy environment. Meanwhile, cheaper hardware is required by the proposed waveforms than their classical counterparts with the same sensing performance. Moreover, flexible resource allocation between the dual functions of the JRC system is supported by the proposed DE-MS-QP waveform, where the optimal parameter design is obtained numerically, enabling accurate target sensing and high-rate communication at the same time.

\begin{figure}[t!]
	\begin{center}
		\includegraphics[width=0.9\linewidth, keepaspectratio]{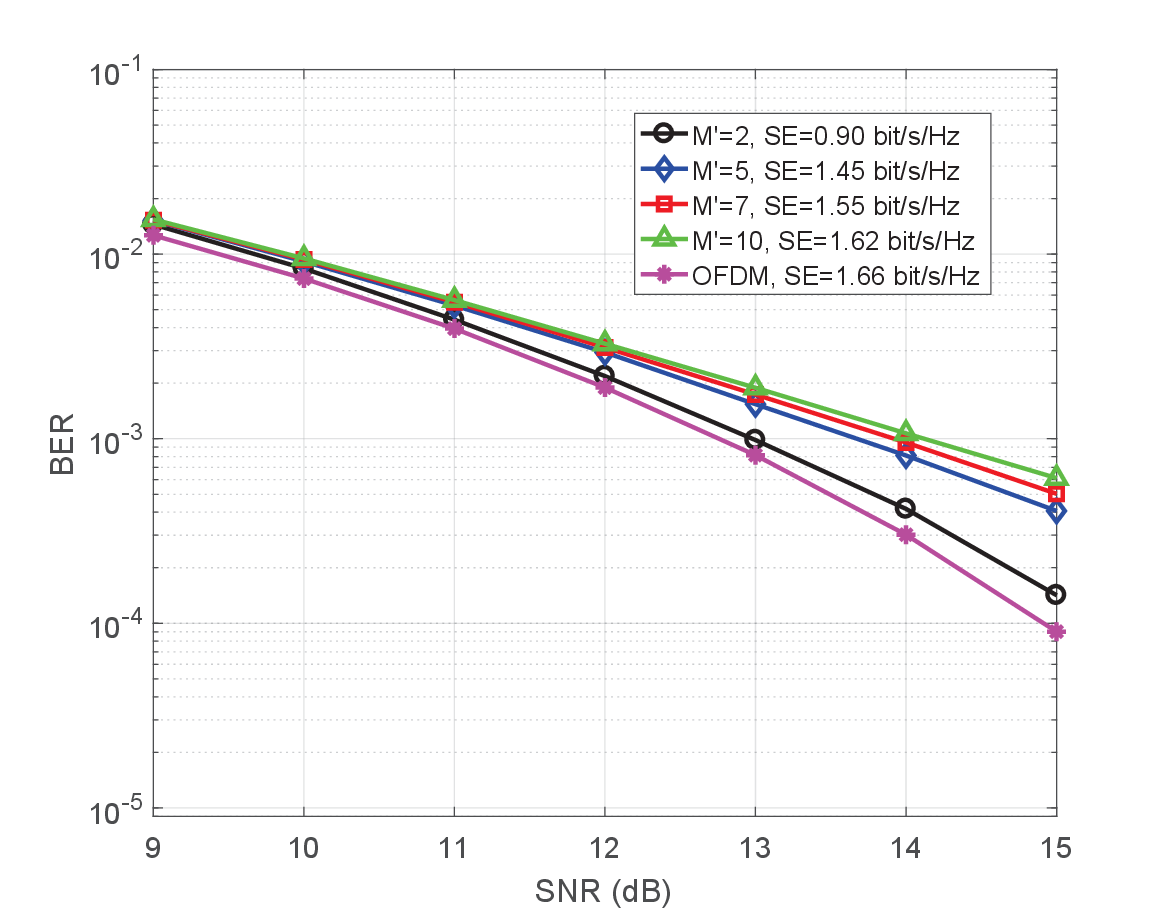}
	\end{center}
	\caption{BER performances of the proposed DE-MS-QP waveforms with different values of $M'$, in comparison with the classical communication waveform.}
	\label{fig15}
\end{figure}

{ There exist open issues for the proposed waveform design. Firstly, without the need of full-band A/D with sampling rate over tens of GHz, the implementation cost of the proposed waveform can be reduced. However, practical use of multiple low-rate A/Ds in parallel could induce synchronization errors of time, frequency and phase. This trade-off between hardware cost and performance loss caused by the synchronization issue should be explored, and hardware/software-based calibration schemes against the synchronization issue of A/Ds also need investigation. Secondly, the values of $N$ and $L_m$ for the MS-QP sequence should be large enough to fight against the severe path loss, which, however, leads to stronger Doppler shift effects. This trade-off in parameter setting of the MS-QP sequence requires further exploration. {Finally, the current waveform design is not robust enough to channel estimation errors and Doppler shifts in terms of communication performance. Whilst imperfect channel estimation is a common issue for wireless communications, the Doppler shift can cause mutual interference between the frequency points within each subband of the proposed DE-MS-QP waveform, leading to difficulty in decoupling the parallel data streams. One possible solution is to integrate the philosophy of differential encoding, which can remove the channel fading coefficient plus the Doppler shift term by differential operations. The in-depth analysis for the impacts of channel estimation errors and Doppler shift will be side aside as our future work. Furthermore, the proposed JRC waveform design will be also enhanced for better communication robustness. }}

\end{document}